\documentclass[a4paper,10pt]{article}
\usepackage{authblk}
\usepackage[top=1.5cm, bottom=2cm]{geometry}
\usepackage{fullpage}
\usepackage[utf8]{inputenc}
\usepackage{cite}
\usepackage{setspace}
\usepackage{algorithm,algorithmic}
\usepackage{xspace}
\usepackage{amsmath,amssymb,amsfonts}
\usepackage{algorithmic}
\usepackage{graphicx}
\usepackage{siunitx}
\usepackage{tabularx}
\usepackage{lineno}
\usepackage{enumitem}
\usepackage{longtable}
\usepackage{booktabs}
\usepackage{newtxtext,newtxmath}
\usepackage{caption}
\usepackage[table]{xcolor}

\title{Future Smart Connected Communities to Fight 
\\ COVID-19 Outbreak}
\author[1]{Deepti Gupta\thanks{deepti.mrt@gmail.com}}
\author[2]{Smriti Bhatt \thanks{sbhatt@tamusa.edu}}
\author[3]{Maanak Gupta \thanks{mgupta@tntech.edu}}
\author[1]{Ali Saman Tosun \thanks{ali.tosun@utsa.edu}}
\affil[1]{Department of Computer Science,
University of Texas at San Antonio,
San Antonio, Texas 78249, USA }
\affil[2]{Department of Computing and Cyber Security, Texas A \& M University-San Antonio,
San Antonio, Texas 78224, USA}
\affil[3]{Department of Computer Science, Tennessee Technological University, Cookeville, Tennessee 38505, USA}
\date{}
\begin{document}
\maketitle
\doublespacing
\begin{abstract}
Internet of Things (IoT) has grown rapidly in the last decade and continue to develop in terms of dimension and complexity, offering wide range of devices to support diverse set of applications. With ubiquitous Internet, connected sensors and actuators, networking and communication technology along with artificial intelligence (AI), smart cyber-physical systems (CPS) provide services rendering assistance and convenience to humans in their daily lives. However, the recent outbreak of COVID-19 (also known as coronavirus) pandemic has exposed and highlighted the limitations of contemporary technological deployments especially to contain the wide spread of this disease. IoT and smart connected technologies together with data-driven applications can play a crucial role not only in the prevention, mitigation or continuous remote monitoring of patients, but also enable prompt enforcement of guidelines, rules and administrative orders to contain such future outbreaks.

In this paper, we envision an IoT and data supported connected ecosystem designed for intelligent monitoring, pro-active prevention and control, and mitigation of COVID-19 and similar epidemics. We propose a gamut of synergistic applications and technology systems for various smart infrastructures including E-Health, smart home, supply chain management, transportation, and city, which will work in convergence to develop `pandemic-proof' future smart communities.
%develop future smart connected communities to manage and mitigate similar outbreaks. 
We also present a generalized cloud-enabled IoT implementation framework along with scientific solutions, which can be adapted and extended to deploy smart connected ecosystem scenarios using widely used Amazon Web Services (AWS) cloud infrastructures.    
%We then present a generalized cloud-enabled IoT implementation architecture utilizing a real-world cloud-IoT platform provided by AWS to implement and deploy smart connected ecosystem scenarios. 
In addition, we also implement an E-Health RPM use case scenario to demonstrate the need and practicality for smart connected communities. Finally, we highlight challenges and research directions which need thoughtful consideration and across the board cooperation among stakeholders to build resilient communities against future pandemics.

% Finally, we discuss challenges and future research directions to enable and develop these smart communities and infrastructures to fight and prepare against such outbreaks.
\end{abstract}

%\begin{keywords}
\textbf{Keywords:} {COVID-19; Coronavirus; Internet of Things; Cloud Computing; Edge Computing; Artificial Intelligence (AI); Machine Learning; Smart Communities; Multi-layered Architecture; Security; Privacy.}
%\end{keywords}

\section{Introduction and Motivation}
\label{sec:introduction}
COVID-19 is an infectious disease caused by a newly discovered coronavirus (SARS-CoV-2) and is rapidly spreading around the world. According to World Health Organization (WHO), COVID-19 has already affected 215\footnote{https://www.who.int/emergencies/diseases/novel-coronavirus-2019} countries and territories around the world and continue to spread rapidly across other regions. The highly contagious coronavirus outbreak was declared a “pandemic” on March 11, 2020. WHO reports\footnote{https://coronavirus.jhu.edu/map.html} that the number of positive cases has dramatically surged, with nearly 34.5 Million reported cases and 1.03 Million fatalities as of October 01, 2020. In order to control the spread the pandemic, lockdowns, quarantines and stay home orders have been issued by several nations across the globe, which have crippled national and world economy with critical consequences to workers, employers and investors. In addition, the industries, businesses and travel restrictions restrain the supply of goods and services, and the economic disruptions will continue to have a long-term impact on global supply chains and economy. In the United States, unemployment rates\footnote{https://www.usnews.com/news/economy/articles/2020-05-08/second-wave-of-coronavirus-joblessness-prompts-comparisons-to-great-depression} spiked to 14.7\% in April 2020 – its highest level since the Great Depression, in addition to fear of stronger second wave of the disease looming during the fall season.

Currently, with no cure or vaccine for this disease, the first line of defense to fight against this pandemic is a combination of preventative measures and mitigation strategies. As suggested by the WHO, the U.S Centers for Disease Control and Prevention (CDC\footnote{https://www.cdc.gov/coronavirus/2019-ncov/index.html}) and several other federal organizations suggest social distancing, environmental surface cleaning, self-isolation, travel restrictions, local and national lockdowns, quarantine, limits on large gatherings, restrictions on opening businesses, and school closures, as some of the preventive measures that are needed to limit the spread of the disease. However, these guidelines impose restrictions which hinder the way of \textit{normal} life for humans. It has become a big challenge to swiftly implement and enforce such measures on a large scale across cities, nations, and around the world. We believe that to effectively enforce and monitor the preventive controls and mitigation strategies for COVID-19, IoT together with its key enabling computing technologies including cloud, AI and data-driven applications can play a pivotal role. 

There are several existing examples of the use of technology to control the spread of COVID-19. Solutions have been proposed to manage large gatherings of individuals to limit the infected cases. The U.S. CDC has introduced a self-checker\footnote{https://www.cdc.gov/coronavirus/2019-ncov/symptoms-testing/testing.html} application enabled with cloud platform, which helps a patient to make decision to find appropriate healthcare service through questionnaires. However, most people do not have any symptoms who are known as the silent spreaders/asymptomatic carriers. Therefore, conducting large scale testing for everyone is paramount to slow the spread. The US is conducting nearly 1 Million\footnote{https://coronavirus.jhu.edu/testing/individual-states} tests everyday (as on Oct 01, 2020) to early determine and isolate patients which also helps in treatment.  Taiwan\footnote{https://jamanetwork.com/journals/jama/article-abstract/2762689} leveraged database from national health insurance and integrated it with its immigration and customs database to begin the creation of big data for analytic, and generated real-time alerts during a clinical visit based on travel history and clinical symptoms to aid case identification. 
%Similarly, Chinese government\footnote{https://www.cnbc.com/2020/03/27/coronavirus-surveillance-used-by-governments-to-fight-pandemic-privacy-concerns.html} installed CCTV cameras at home or apartment door of those under a quarantine period to monitor the patient activities. Drones are used to notify people to wear their masks, and digital barcodes on mobile apps highlight the health status of individuals. 
%Researchers at the University of Michigan together with the team at Voxel51\footnote{https://pdi.voxel51.com/} have presented the Physical Distancing Index (PDI) to help track how COVID-19 news or events impact human activity around the globe in real time. The free AI powered interactive tool enables users to explore a day-by-day timeline of social activity in some of the world’s most populated areas to track social distancing behaviors. Currently, Voxel51 gathers and analyzes historical and real time video streams from public street cameras in  Ann Arbor, Detroit, Times Square; Abbey Road in London; Fremont Street in downtown Las Vegas; Seaside Heights in New Jersey; a beach in Ft. Lauderdale; and intersections in Dublin and Prague.

In April, New York City\footnote{https://www1.nyc.gov/site/doh/covid/covid-19-data.page} reported 13\% positive cases of COVID-19. The fatality rate of this city is 11\% as of May 10, and 0.5\% of fatalities have been in nursing homes. Nursing homes are the most vulnerable spot in terms of health risks. Technology can assist in enforcing the prevention measures and mitigation strategies in nursing homes to flatten the curve of fatality rate. Some states Iowa, Minnesota, Tennessee, and Texas have partially reopened, and the models\footnote{https://www.nytimes.com/2020/05/04/us/coronavirus-live-updates.html} have predicted that the number of cases and fatalities will increase as the country moves towards reopening businesses and cities.
These evidences show that communities are not prepared to reopen, operate, and handle such pandemics. There are several requirements to assist in tracking and monitoring COVID-19 patients, such as to enforce wearing masks in public area, proper sanitizing, social distancing, mass quarantine for mild-symptoms people, and testing at large scale. According to Harvard Global Health Institute\footnote{https://globalepidemics.org/2020/04/27/states-that-fall-short-on-testing/}, 31 states in United States have insufficient testing levels, and 10 states would need at least 10,000 more tests a day to begin a gradual reopening. For instance, in April 2020, New York, the state with the largest number of cases, needed to increase testing by more than 10,000 per day. Due to lack of enough testing kits, an antibody test online is offered for those people who recently got sick and think that they might have COVID-19. 

%Over the last few years, there has been a huge surge in the number of IoT devices and different types of smart sensors offering variety of services. 
In today's world, not having network capability in a device limits the market potential for that device. As a result, there are large number and various types of network connected IoT devices providing convenience and ease of life to humans. With new technological advancements, this trend is expected to continue and grow in the future. IoT market\footnote{https://www.forbes.com/sites/louiscolumbus/2018/08/16/iot-market-predicted-to-double-by-2021-reaching-520b/\#82674f91f948} is currently valued at \$267 billion per year and is expected to reach \$520 billion by 2021. Another recent article\footnote{https://www.cisco.com/c/dam/en/us/products/collateral/se/internet-of-things/at-a-glance-c45-731471.pdf} predicted more than 100 billion devices to be internet-connected by 2025.
\begin{figure}[t]
\centering
\includegraphics[width=.8\textwidth]{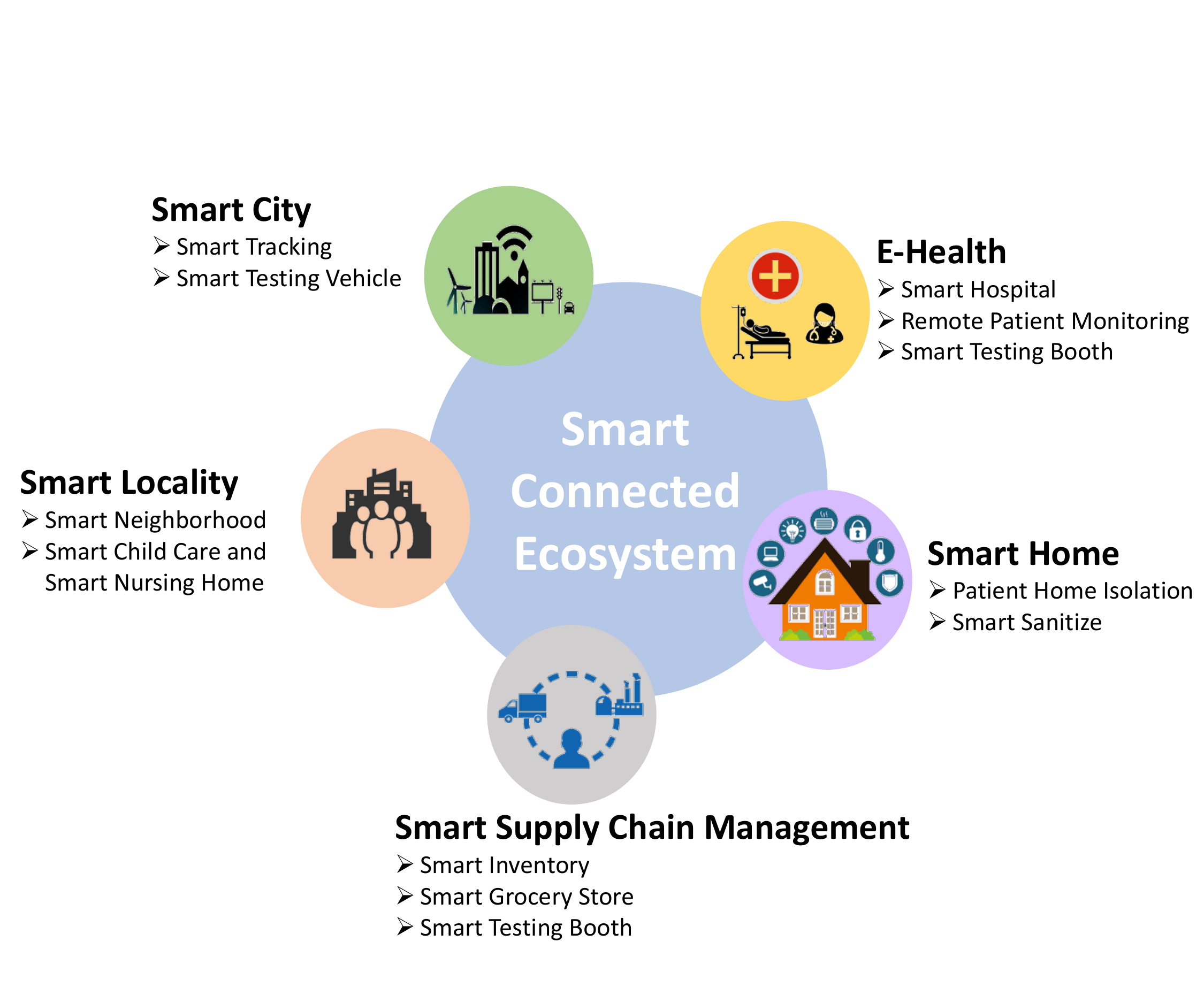}
\centering
\caption{An Overview of Converging Smart Connected Ecosystem}
\label{fig:usecases}
\end{figure}
IoT comprises large number of novel consumer devices including HDMI sticks, IP cameras, smartwatches, connected light bulbs, smart thermostats, health and fitness trackers, smart locks, connected sprinkler systems, garage connectivity kits, window and door sensors, smart light switch, home security systems, smart ovens, smart baby monitors, and blood pressure monitors. However, mostly these IoT devices are used in a distributed manner based on users' requirements. IoT including smart sensors, actuators, and devices and data driven applications can enable future smart connected communities to strengthen the health and economical postures of the nations to fight against the current COVID-19 situation and other future pandemics efficiently. 
These smart devices have the potential to be a major breakthrough in efforts to control and fight against the current pandemic situation. IoT is an emerging field of research, along with the ubiquitous availability of smart technologies, as well as increased risks of infectious disease spread through the globalization and interconnection of the world necessitates its use for predicting, monitoring and tracking to prevent from COVID-19.

%We have been wondering to what extent IoT technology can be used for Covid-19 outbreak. Individual pieces of the puzzle were either available or developed in the last few months. 
In this paper, we present a holistic vision of IoT-enabled smart communities utilizing various IoT devices, applications, and relevant technologies (e.g., AI, Machine Learning (ML), Blockchain etc.). Figure \ref{fig:usecases} shows an overview of smart connected ecosystem with real-world scenarios in diverse applications domains focusing on detection, prevention and mitigation of COVID-19 and similar outbreaks. The major contributions of this paper are as follows.
\begin{itemize}
\item We present an extensive review of IoT along with emerging technologies currently used (and proposed) to control pandemic crisis, and enlists the focus, contributions and weaknesses of relevant research works. We outline some of COVID-19 symptoms, preventive measures, mitigation strategies, and current problems, challenges to understand the needs of proposed connected ecosystem.
\item We present an overview of adaptable multi-layered IoT architecture supporting edge, virtual objects and cloud layer, and reflect interactions between them to focus on data driven and connected smart applications catering to limit the COVID-19 outbreak. 
\item We design a smart connected ecosystem by developing multiple synergistic IoT application frameworks including E-Health, Smart Home, Supply Chain Management, Transportation, and Smart City. We introduce the convergence of these domains and offer novel use cases and application scenarios for early COVID-19 detection, prevention and mitigation.
\item We present a general implementation framework that can be used for adapted and extended to enable smart communities use case scenarios utilizing a real-world cloud-enabled IoT platform, Amazon Web Services (AWS). We also design specific algorithms which can be used in different scenarios as required. In addition, we also implement a Remote Patient Monitoring (RPM) use case within E-Health domain very relevant to COVID-19 patients in home isolation and enforcing quarantine. 
\item Lastly, we identify and highlight current challenges and interdisciplinary research directions, including security and privacy, performance efficiency, interoperability and IoT federation, implementation challenges, policy and guidelines, ML and big data analytics, to enable and empower future smart connected communities. 
\end{itemize}

The remainder of this paper is organized as follows. Section \ref{related} presents the literature review on role of computing and data driven technologies with IoT to control COVID-19. Section \ref{sec:characteristics} discusses the essential characteristics to diagnose, prevent and mitigate COVID-19 disease. Section \ref{sec:model} presents a multi-layered architecture for IoT enabled future smart and resilient communities, whereas Section \ref{sec:usecases} discusses convergence of connected ecosystem scenarios in various IoT application domains. Section \ref{sec:implementation} discusses a general implementation framework of smart connected communities using real-world cloud-enabled IoT platform and deploys a proof of concept test bed for Remote Patient Monitoring (RPM) use case. Section \ref{sec:challenges} highlights open research challenges and future directions, followed by conclusion in Section \ref{sec:conclusion}.

\section{Related Work and Relevant Background Study}
\label{related}
Extending the use of IoT along with emerging technologies can play a significant role in dealing with COVID-19. This work justifies that IoT has influenced healthcare and other areas of our lives significantly. In current situation, hospital practitioners are working hard to help infected patients, and we are living a new normal with new rules, such as state leaders' mandate rules to maintain six feet distance and wear masks in public places. Broaden use of IoT technology enables us to deal with COVID-19 or any future pandemic situation. For instance, Chinese government\footnote{https://www.cnbc.com/2020/03/27/coronavirus-surveillance-used-by-governments-to-fight-pandemic-privacy-concerns.html} installed CCTV cameras at home or apartment door of those under a quarantine period to monitor the patient activity. Human activities also measured through a linear model \cite{kabir2019state} based on the state-space method. Drones are used to notify people to wear their masks, and digital barcodes on mobile apps highlight the health status of individuals. Researchers at the University of Michigan together with the team at Voxel51\footnote{https://pdi.voxel51.com/} have presented the Physical Distancing Index (PDI) to help track how COVID-19 news or events impact human activity around the globe in real-time. The free AI powered interactive tool enables users to explore a day-by-day timeline of social activity in some of the world’s most populated areas to track social distancing. Currently, Voxel51 gathers and analyzes historical and real-time video streams from public street cameras in Ann Arbor, Detroit, Times Square; Abbey Road in London; Fremont Street in downtown Las Vegas; Seaside Heights in New Jersey; a beach in Ft. Lauderdale; and intersections in Dublin and Prague.

Table \ref{tab:1} summarizes prior research along with their challenges and contributions to prevent and mitigate COVID-19 using specific technologies, such as IoT, ML, and Blockchain. In existing literature \cite{bai2020chinese, vaishya2020artificial, javaid2020industry, kumar2020role, vinod2020data, otoom2020iot, kumar2020drone, lalmuanawma2020applications, singh2020internet, swayamsiddha2020application, elavarasan2020restructured, vafea2020emerging, alam2020internet, albahri2020role}, various IoT-based solutions along with other technologies are proposed for smart healthcare. Various frameworks and systems are discussed in the context of healthcare for handling COVID-19. Here, critical analysis of current IoT based research shows that most of them are focused on a single use case; however, we need holistic smart ecosystem comprising multiple entities and different stakeholders to handle current or similar future pandemic situations. Therefore, we present future smart connected communities to fight COVID-19 and appropriately handle future pandemic situations.
\begin{center}
\begin{spacing}{0.5}
\renewcommand{\arraystretch}{2.5}
\begin{longtable}{p{3cm}|p{3cm}|p{5cm}|p{4cm}}
\caption{Current projects and research on IoT based solutions to mitigate COVID-19.} \label{tab:1} \\
\hline \multicolumn{1}{c|}{\textbf{Paper Title}} & \multicolumn{1}{c|}{\textbf{Focus/Objective}} & \multicolumn{1}{c|}{\textbf{Contribution}} & \multicolumn{1}{c}{\textbf{Limitations}} \\ \hline 
\endfirsthead

\multicolumn{4}{c}
{{ \tablename\ \thetable{}. (Continued.) Current projects and research on IoT based solutions to mitigate COVID-19.}} \\
\hline \multicolumn{1}{c|}{\textbf{Paper Title}} & \multicolumn{1}{c|}{\textbf{Focus/Objective}} & \multicolumn{1}{c|}{\textbf{Contribution}} & 
\multicolumn{1}{c}{\textbf{Limitations}}\\ \hline 
\endhead

\hline 
%\multicolumn{3}{|r|}{{Continued on next page}} \\ \hline
\endfoot

\hline \hline
\endlastfoot \rowcolor[gray]{.9}
%\textbf{Paper Title}  & \textbf{Focus/Objective} & \textbf{Contribution}  &\textbf{Limitations}\\\rowcolor[gray]{.9} \hline
Chinese experts’ consensus on the Internet of Things-aided diagnosis and treatment of coronavirus disease 2019 (COVID-19)\cite{bai2020chinese} 
& Enables different levels of COVID-19 diagnosis and treatment among different doctors from different hospitals through the intelligent assistance of the nCapp system. 
& 

\begin{itemize}[leftmargin=*,topsep=0pt,parsep=0pt]
\item Develops COVID-19 Intelligent Diagnosis and Treatment Assistant Program (nCapp) based on the Internet
of Things.
\item Designs 15 easy-to-use questionnaires for intelligent processing.

\end{itemize}
& \begin{itemize}[leftmargin=*,topsep=0pt,parsep=0pt]
\item The data is based on questionnaires and is filled manually. 
\item Security and privacy issues related to patient's personal data have not been discussed.
\end{itemize}
\\
Artificial Intelligence (AI) applications for COVID-19 pandemic\cite{vaishya2020artificial}
& The role of AI as a decisive technology to analyze and prepare us for prevention and fight with COVID-19
& \begin{itemize}[leftmargin=*,topsep=0pt,parsep=0pt] 
\item Identifies seven significant AI assisted applications for fight with COVID-19. 
\item Detects the cluster of positive cases and predicts where this virus will affect in future by collecting and analyzing all previous data.
\end{itemize}
& \begin{itemize}[leftmargin=*,topsep=0pt,parsep=0pt] 
\item Tools are missing to implement each application. The answer of "How will author extract the data?" is missing. 
\item The flow of each application has not been explained.
\end{itemize}
\\\rowcolor[gray]{.9}
Industry 4.0 technologies and their applications in fighting COVID-19 pandemic\cite{javaid2020industry}
& Explores several useful technologies of Industry 4.0 to control and manage of COVID-19 pandemic. 
& \begin{itemize}[leftmargin=*,topsep=0pt,parsep=0pt] 
\item Presents ten technologies- Artificial intelligence, Big data, Internet of Things, Virtual reality, Holography, Cloud
computing, Autonomous robot, 3D Scanning, 3D Printing and Biosensor. 
\end{itemize}
& \begin{itemize}[leftmargin=*,topsep=0pt,parsep=0pt] 
\item Only medical concern has been discussed here, transportation and other topics related to industry have not been discussed.  
\end{itemize}
\\
Role of IoT to avoid spreading of COVID-19\cite{kumar2020role}
& Presents available literature on COVID-19 regarding monitoring techniques, and suggests an IoT based architecture, which can be used to minimize the spreading of COVID-19.
& \begin{itemize}[leftmargin=*,topsep=0pt,parsep=0pt] 
\item Presents IoT architecture to avoid COVID-19, the use of IoT with smart sensors to measure and record the body temperature of individuals, which can help to identify infected individuals.
\end{itemize}
& \begin{itemize}[leftmargin=*,topsep=0pt,parsep=0pt] 
\item The number of sensors to develop particular application are limited. 
\item Only one application is used for per scenario, complete framework is missing. 
\item The upcoming challenges to build these applications have not been discussed here. 
\end{itemize}
\\\rowcolor[gray]{.9}
Data science and the role of Artificial Intelligence in achieving the fast diagnosis of Covid-19\cite{vinod2020data}
& Due to less number of COVID-19 test kits, presents image classification approach on chest X-ray to predict COVID-19 test result. 
& \begin{itemize}[leftmargin=*,topsep=0pt,parsep=0pt]
\item Elaborates a methodology that helps identify COVID-19 infected people among the individuals by utilizing CT scan and chest X-ray images using AI. 
\item Experiment on a dataset of COVID-19 and normal chest X-ray images. 
\item Calculates accuracy of an image using decision tree classifier. 
\end{itemize}
& \begin{itemize}[leftmargin=*,topsep=0pt,parsep=0pt]
\item False positive rate can be high in allergy case.
\item The data is not categorized by age. 
\item No discussion on security and privacy challenges to collect and store the patients' health data.
\end{itemize}
\\

An IoT-based Framework for Early Identification and Monitoring of COVID-19 Cases\cite{otoom2020iot}
& Identification of potential cases of COVID-19. 
& \begin{itemize}[leftmargin=*,topsep=0pt,parsep=0pt]
\item Presents a novel IoT framework, which consists of five major components. 
\item Uses eight classification algorithms on a real COVID-19 symptom dataset using IoT framework to identify COVID-19 cases. 
\end{itemize}
& \begin{itemize}[leftmargin=*,topsep=0pt,parsep=0pt]
\item The generated data can be bias, if number of COVID-19 cases are less than other type of diseases then the model can send false alarm. 
\end{itemize}
\\\rowcolor[gray]{.9}
Internet of Things and Blockchain-based framework for Coronavirus (Covid-19) Disease.\cite{alam2020internet}
& Aims to assist infected people online using the
Internet of Things (IoT) and Blockchain technologies through smart devices. 
&\begin{itemize}[leftmargin=*,topsep=0pt,parsep=0pt]
\item Presents four-layer architecture: hardware layer, shared ladger layer, communication layer, Application layer.
\item Develops IoT and Blockchain-based framework for healthcare.
\end{itemize}
&\begin{itemize}[leftmargin=*,topsep=0pt,parsep=0pt]
\item Raise security concerns with shared ladger layer to store health records of patient. 
\item Challenges to build the framework have not been explained. 
\end{itemize}
\\
A drone-based networked system and methods for combating coronavirus disease (COVID-19) pandemic\cite{kumar2020drone}
& Investigates drone-based systems, which are used to handle pandemic situations in real-world scenarios.
&\begin{itemize}[leftmargin=*,topsep=0pt,parsep=0pt]
\item Presents an artificial intelligence-based system that collects data through drones. 
\item Develops a multi-layered architecture that collects information from drones.
\item Implements a real-time drone-based system for sanitization, monitoring, and other applications.
\end{itemize}
&\begin{itemize}[leftmargin=*,topsep=0pt,parsep=0pt]
\item No user's data privacy solutions.
\item Limited scope on data security. 
\end{itemize}
\\\rowcolor[gray]{.9}

Applications of machine learning and artificial intelligence for Covid-19 (SARS-CoV-2) pandemic: A review\cite{lalmuanawma2020applications}
& Presents review of AI and ML as one significant method in the arena of screening, predicting, forecasting, contact tracing, and drug development for SARS-CoV-2 and its related epidemic.

& \begin{itemize}[leftmargin=*,topsep=0pt,parsep=0pt]
\item Discusses recent studies that apply ML and AI technologies towards augmenting current research works from multiple angles. 
\item This study also addresses a few errors and challenges while using such algorithms in real-world problems. 
\end{itemize}
&\begin{itemize}[leftmargin=*,topsep=0pt,parsep=0pt]
\item AI and ML based algorithms have not been discussed. 
\item A complete framework is not shown, which can bring all small applications together. 
\end{itemize}
\\%\rowcolor[gray]{.9}
Internet of things (IoT) applications to fight against COVID-19 pandemic\cite{singh2020internet}
& Enables Internet of Things (IoT) based healthcare system, which helps to increase patient satisfaction and reduces re-admission rate in the hospital. 
& \begin{itemize}[leftmargin=*,topsep=0pt,parsep=0pt]
\item IoT provides an extensive integrated network for healthcare providers to fight with COVID-19 pandemic.
\item Presents twelve major applications of IoT for COVID-19 pandemic. 
\end{itemize}
&\begin{itemize}[leftmargin=*,topsep=0pt,parsep=0pt]
\item Any healthcare related IoT device has not been presented.
\item A novel implementation of any application has not been described. 
\item Challenges to design these twelve applications are also missing in this paper. 
\end{itemize}
\\\rowcolor[gray]{.9}
Application of cognitive Internet of Medical Things for COVID-19 pandemic\cite{swayamsiddha2020application}
& Cognitive Internet of Medical Things (CIoMT) is best suited to handle COVID-19.
&\begin{itemize}[leftmargin=*,topsep=0pt,parsep=0pt]
\item Presents the CIoMT platform, which enables real-time
tracking, remote health monitoring and other applications to reduce the workload on the medical industry.
\item Identifies some challenges and future research directions.
\end{itemize}
&\begin{itemize}[leftmargin=*,topsep=0pt,parsep=0pt]
\item No implementation details of any application is discussed.
\item The details of other major challenges like interoperability, data biasing are missing. 
\end{itemize}
\\%\rowcolor[gray]{.9}
Restructured society and environment: A review on potential technological strategies to control the COVID-19 pandemic\cite{elavarasan2020restructured}
& Investigates various implemented technologies that assists the healthcare systems, government and public in diverse aspects for fighting against COVID-19.
& \begin{itemize}[leftmargin=*,topsep=0pt,parsep=0pt]
\item Presents prospective viable technologies that can be used in the current or future epidemic situation.
\item Discusses the technological changes that the environment and the society have undergone in tackling COVID-19.
%\item Suggests few drafted pioneering systems in mitigating the spread.
\end{itemize} 
&\begin{itemize}[leftmargin=*,topsep=0pt,parsep=0pt]
\item Limited technological solutions for COVID-19 risk.
\item Based on the survey, any novel idea is not presented. 
\end{itemize}
\\\rowcolor[gray]{.9}
Emerging Technologies for Use in the Study, Diagnosis, and Treatment of Patients with COVID-19.\cite{vafea2020emerging}
& Focuses on analyzing possible opportunities and
challenges of integrating emerging technologies into COVID-19 contact tracing.
&\begin{itemize}[leftmargin=*,topsep=0pt,parsep=0pt]
\item Designs the contact tracing application using emerging technologies to mitigate the coronavirus. 
\item Observes various challenges like the security and privacy of people while using contact tracing application. 
\end{itemize}
&\begin{itemize}[leftmargin=*,topsep=0pt,parsep=0pt]
\item It is not clear how effective the current contact tracing application would be using emerging technologies. 
\item Lack of implementation details.
\end{itemize}
\\%\rowcolor[gray]{.9}
Role of biological Data Mining and Machine Learning Techniques in Detecting and Diagnosing the Novel Coronavirus (COVID-19): A Systematic Review\cite{albahri2020role}
& Aims to address the limitations of utilizing data mining and ML algorithms to mitigate this virus, and provides health sector with the benefits of techniques.
&\begin{itemize}[leftmargin=*,topsep=0pt,parsep=0pt]
\item Reviews eight articles, which show data mining and ML based applications. 
\item Analyzes several database, and performs various AI algorithms for detection, and classification of adaptive CoV.
\end{itemize}
&\begin{itemize}[leftmargin=*,topsep=0pt,parsep=0pt]
\item Analyzes only MERS, SARS disease datasets (2013-2016). 
\end{itemize}
\\\rowcolor[gray]{.9}
\hline
%\end{tabular}
\end{longtable}
\end{spacing}
\end{center}

\section{Essential Characteristics to Diagnose, Prevent and Mitigate COVID-19}
\label{sec:characteristics}
Coronavirus transmits mainly by the infected person's saliva and nasal drips which spread during coughing and sneezing around anybody in close contact. Another source of infection is contaminated surfaces in surrounding and high-risk areas, such as door handles, railings, elevators, and public restrooms. COVID-19 is a highly contagious virus with the incubation period stretched from 2 days to 2 weeks after exposure. Symptoms of COVID-19 range from mild symptoms including fever, coughing and shortness of breath to severe symptoms including organ failure, such as kidney failure, and pneumonia. Researchers are still studying the virus to fully understand the characteristics and new symptoms of this disease. Recently, the U.S. CDC\footnote{https://www.cdc.gov/coronavirus/2019-ncov/symptoms-testing/symptoms.html} added six new possible symptoms including chills, muscle pain, headache, sore throat and new loss of taste or smell. New York City\footnote{https://www.nbcnewyork.com/news/local/73-ny-children-sick-with-rare-covid-related-illness-state-finds-with-one-death/2408285/} reported 73 cases of children with rare coronavirus inflammatory illness on May 7, 2020.

At this time, any specific treatment is not recommended for disease caused by the novel coronavirus, and no vaccine is currently available in the market. The treatment is subjective and depends on a case-by-case basis, where oxygen therapy shows positive affects as the treatment intervention for patients with severe infection. Mechanical ventilation may be necessary in cases of respiratory failure refractory to oxygen therapy. The U.S.  CDC\footnote{https://www.cdc.gov/coronavirus/2019-nCoV/index.html} and other resources explained early signs and symptoms, some preventive measures, and mitigation strategies. On March 19, 2020, the WHO\footnote{https://www.who.int/publications-detail/infection-prevention-and-control-during-health-care-when-novel-coronavirus-(ncov)-infection-is-suspected-20200125} released the first edition of interim guidance on Infection Prevention and Control (IPC) strategies for identifying the coronavirus infection.

On March 28, 2020, the U.S. Food and Drug Administration (FDA\footnote{https://www.fda.gov/media/136537/download}) provided emergency use authorization for hydroxychloroquine medicine to treat the people who are suffering from this virus in hospitals. Later, on April 24, 2020 FDA\footnote{https://www.fda.gov/drugs/drug-safety-and-availability/fda-cautions-against-use-hydroxychloroquine-or-chloroquine-COVID-19-outside-hospital-setting-or} warned against use of hydroxychloroquine to treat this disease outside of the hospital setting or a clinical trial due to risk of heart rhythm problems. 
People can protect themselves by following some protective measures and help to slow the spread using mitigation strategies. Table \ref{tab:2} provides a comprehensive overview of the symptoms, preventive measures, mitigation strategies and some challenges fighting COVID-19 disease. 

One of the easiest preventive measure is to wash your hands frequently and thoroughly with soap and water for at least 20 seconds or use hand sanitizer or an alcohol-based hand rub when soap and water are not available. People should keep social distancing (six feet distance) from others especially from people who are coughing or sneezing. It is suggested to wear mask and gloves in public setting, and avoid touching the face and surfaces such as the button at a traffic light, a keypad to add a tip for the restaurant take-out order, elevator buttons, etc. Many surfaces are touched by hands accidentally and virus can be potentially picked up and then transmitted to other surfaces and locations. Once the hands are contaminated, the virus can be transferred through eyes, nose or mouth, thus, it enters human body. Respiratory Hygiene/Cough Etiquette is a term used to describe infection prevention measures to decrease the transmission of virus. There is a need to avoid cough or sneeze into the hands, and to cover mouth and nose with a tissue and throw away the tissue immediately. Groceries and packets can be contaminated from coronavirus, and it is recommended to wash grocery items carefully and wipe packets using disinfectant spray. Local public health administrations regularly issue health guidelines, which people should follow.
\begin{center}
\begin{spacing}{0.5}
\renewcommand{\arraystretch}{2.5}
\begin{longtable}{p{3.5cm}|p{3.5cm}|p{3.5cm}|p{3.5cm}}
\caption{COVID-19 Symptoms\protect \footnote{https://www.cdc.gov/coronavirus/2019-ncov/symptoms-testing/symptoms.html}, Prevention\protect\footnote{https://www.cdc.gov/coronavirus/2019-ncov/prevent-getting-sick/prevention.html}, Mitigation and Challenges. \\ (Items within the same row are unrelated)} \label{tab:2}\\
\hline \multicolumn{1}{c|}{\textbf{Early Symptoms}} & \multicolumn{1}{c|}{\textbf{Preventive Measures}} & \multicolumn{1}{c|}{\textbf{Mitigation Strategies}} & \multicolumn{1}{c}{\textbf{Problems and Challenges}} \\ \hline
\rowcolor[gray]{.9}
Common symptoms include fever, dry cough and myalgia or fatigue. 
& 
Clean hands often for at least 20 seconds with soap and water or use of alcohol-based hand sanitizer.
&
If sick stay in a single room for 14 days, avoid sharing personal household items.
& 
Coping with anxiety disorder, depression issues, and mental health problems.\\ 
Shortness of breath or cannot breathe deeply enough to fill your lungs with air, chills. 
& 
Avoid face-to-face meetings, practice social distance from other people outside of the home. 
& 
Monitor symptoms regularly, wear a cloth covering or N-95 mask over nose and mouth.
&
Knowledge gaps to understand virus transmission, no specific antiviral treatment, and no vaccine available.  \\ \rowcolor[gray]{.9} 
Loss of the sense of smell is most likely to occur by the third day of infection and some patients also have experience a loss of the sense of taste.
& 
Cover mouth and nose with a cloth or wear mask when around others, wear gloves and discard them properly.
& 
Manufacturers use of all cleaning and
disinfection products, follow the workplace protocol\footnote{https://www.osha.gov/Publications/OSHA3990.pdf} and provide PPE to their employee. 
& 
Lack of testing and essential resources such as ventilators, masks, beds, and health staffs, cancel elective surgery. \\ 
Diarrhea and nausea a few days prior to fever, sudden confusion or an inability to wake up and be alert may be a serious sign. 
& 
Cover your mouth and nose with a tissue when you cough or sneeze; Throw used tissues in the trash.
&
Hospital task force such as increase the number of testing, available the PPE for their staff members, and increase the incentive care.
&
Privacy issues: contact tracing, health data/medical records, virtual meeting, remote workforce.\\\rowcolor[gray]{.9} 
A small number of patients can have headache or hemoptysis and even relatively asymptomatic.
& 
Maintain proper hygiene, clean and disinfect frequently touched surfaces, and wash grocery items properly.
& 
City/state government task force such as limited to 25\% capacity at retail/restaurants, close the playground, and restrict or limit visitor access to nursing homes.  
& 

Reopening the country phases in highly contagious environment without data analytic, unified decision-making frameworks and some policies that span the country.\\ 

Pneumonia, kidney failure and dyspnea more frequent in most severe cases. 
&
Follow travel restrictions (domestic flights only for essential services, No sail order) &
Individuals limit community movement and adapt to disruptions in routine activities (school and/or work closures) according to guidance from local officials.
&
Due to novelty of the virus, projection of the model is unpredictable to identify number of positive cases and fatalities, and evolving symptoms such as COVID toes and rashes. \\\rowcolor[gray]{.9}
Normal or low white blood cell count or reduced lymphocyte in early onset. &
Update periodically to follow WHO, and country guidelines.
&
Use existing technologies to mitigate the risk of virus.  & 
No strict and defined guidelines from agencies due to ever changing dynamics of the virus.
\\ \hline
\end{longtable}
\end{spacing}
\end{center}
The WHO, governments, and healthcare workers are all urging people to stay home if they can. On top of basic illness prevention, experts said that the best (and only real) defense against disease is a strong immune system. In addition to the physical health, taking care of mental health is also necessary. High stress levels can take a toll on human's immune system, which is the opposite of what people want in this situation. 
In addition, mitigation strategies are a set of actions applied by the people and communities (hospitals, grocery stores, and cities) to help slow the spread of respiratory virus infections. These actions can be scaled up or down depending on the evolving local situation. At individual level, if a person is infected with coronavirus, then he/she should self-isolate and follow the guidelines of quarantine provided by hospital. The hospitals must support healthcare workforce, increase testing and intensive care capacity, and availability of Personal Protective Equipment (PPE). City governments can appoint task force, open shelters for homeless people, and maintain availability of resources to implement preventive and mitigation strategies for this disease. While each community is unique, appropriate mitigation strategies vary based on the level of community transmission, characteristics of the community and their populations, and the local capacity to implement strategies. Nonetheless, it is crucial to understand the characteristics of this novel virus, spread awareness and up-to-date information across communities through appropriate technologies. Consequently, it is essential to address the challenges with significant research and implementation of strategies as shown in the Table \ref{tab:2}.
\begin{figure}[t]
\centering
\includegraphics[width=.7\textwidth, height = .35\textheight]{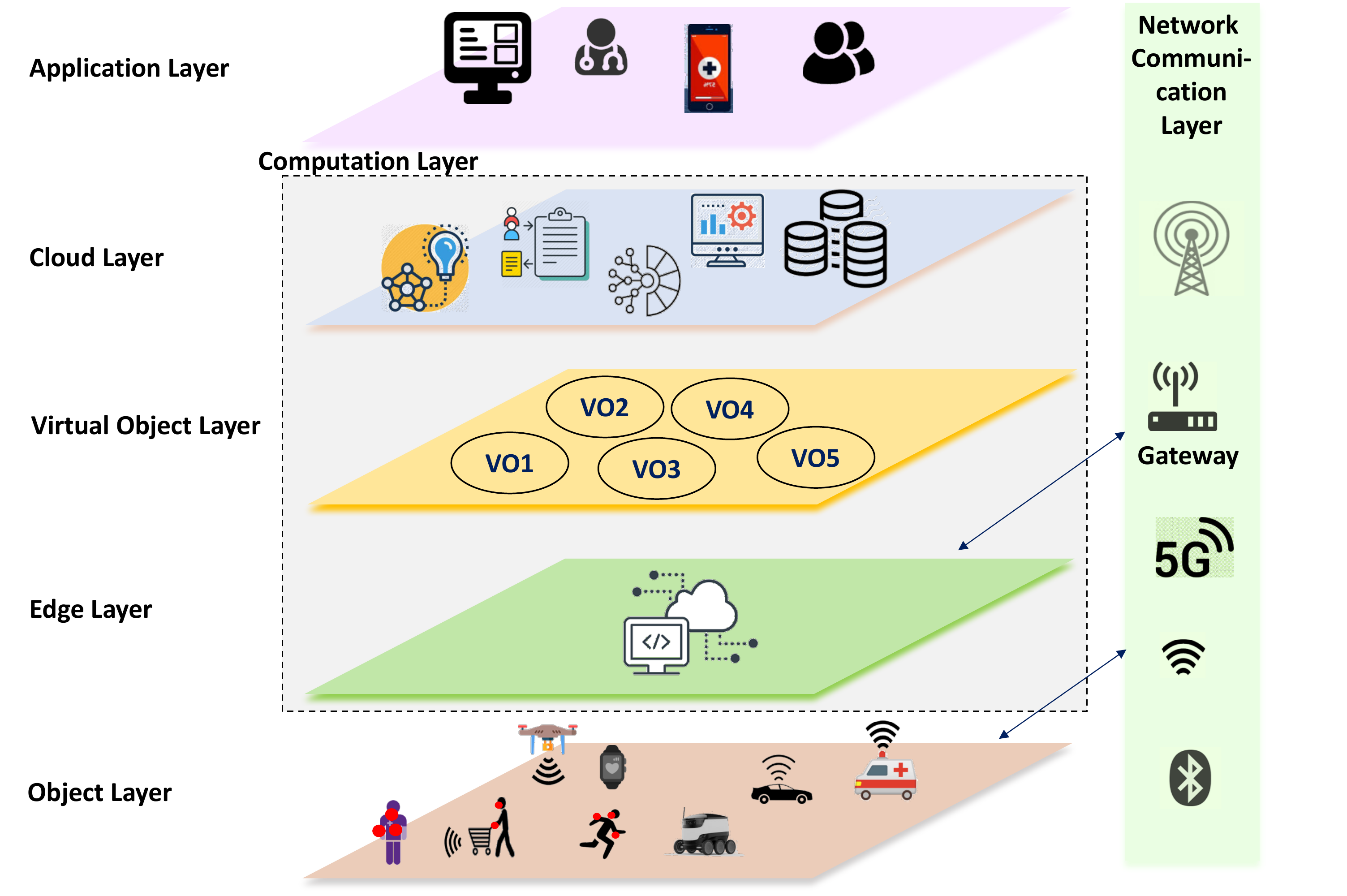}
\centering
\caption{Adaptable Multi-Layered Smart Communities Architecture.}
\label{fig:arch}
\end{figure}

\section{Multi-Layered Smart Communities Architecture}%
\label{sec:model}
In this section, we explain an integrated multi-layer IoT architecture which can fundamentally change the infrastructure and underlying technologies for smart communities including hospital, grocery retail store, transportation, and city etc. as shown in Figure \ref{fig:arch}. Our proposed architecture extends and adapts echinaxisting IoT and CPS architectures \cite{stantchev2015smart, li2014enabling, fernandez2014opportunities, gupta2020security, hossain2016cloud, bhatt2017access,yadav2019docker}, and focuses on the need of swift enforcement of policies, laws and public guidelines, in order to curtail the widespread of such disease. The architecture integrates a hybrid cloud and edge computing nodes together with IoT and smart sensor devices, to enable real-time and data-driven services and applications needed in COVID-19 pandemic. Overall, the architecture consists of six layers: Object layer, Edge layer, Virtual Object layer, Cloud layer, Network Communication, and Application layer. The object layer is a rich set of IoT devices including sensors, actuators, embedded devices, road side infrastructures, vehicles, etc. These physical objects are spread across and implemented in smart communities, such as hospitals, retail-stores, homes, parking lots. The edge layer provides local real-time computation and analysis needed for smart resource constrained physical objects. This layer incorporates edge gateways and cloudlets \cite{satyanarayanan2009case} which can enable local computation at this layer overcoming limited bandwidth and latency requirements, and also impacts the usability of the IoT applications.

This multi-layer architecture has integrated the concept of Virtual Objects (VOs) \cite{nitti2015virtual}, which are the digital delineation of physical IoT devices. VOs show the current state of corresponding physical objects in the digital space when they are connected, and can also store a future state for these devices when they are offline. The cloud layer provides various services like remote storage, computation, big data analysis, and data-driven AI applications etc. for huge amount of information generated by billions of IoT devices connected to the cloud. We define a computation layer which comprises of edge layer, virtual object layer, and cloud layer. Computation, data analytic and processing services are performed in this layer. Network communication layer run among different layers to establish the interaction. It is responsible for connecting physical sensors, smart devices, edge compute nodes or cloudlets, and cloud services with different technologies, and is also used for transmitting and processing sensor data. Application layer delivers specific services to end users through different IoT applications. Task scheduling algorithms \cite{pandey2019application} can be applied on this layer to manage the resources in order to deliver the optimal Quality of Services (QoS). In the multi-layered smart communities' architecture, this application integrates mobile phones, edge computing, cloud computing, AI based analytic, and data-driven services.

This architecture can incorporate the IoT application frameworks within different domains as discussed in the Section \ref{sec:usecases}, and different use case scenarios can be mapped and implemented using relevant technologies associated with each layer of the architecture.

\section{Future Smart Connected Ecosystem and Technological Solutions}
%Smart Connected Ecosystem Scenarios
\label{sec:usecases}
%\textcolor{red}{Need to change the section heading}
We propose IoT use case scenarios by deploying novel smart devices, data-driven applications and technologies to present a holistic view of IoT-enabled smart architectures for fighting COVID-19 outbreaks. We have divided our scenarios into five broader categories: E-Health, Smart Home, Smart Supply Chain, Smart Locality and Smart City, as described in the following subsections.

\subsection{E-Health} 
\begin{figure*}[t]
\centering
\includegraphics[width=1\textwidth, height=.4\textheight]{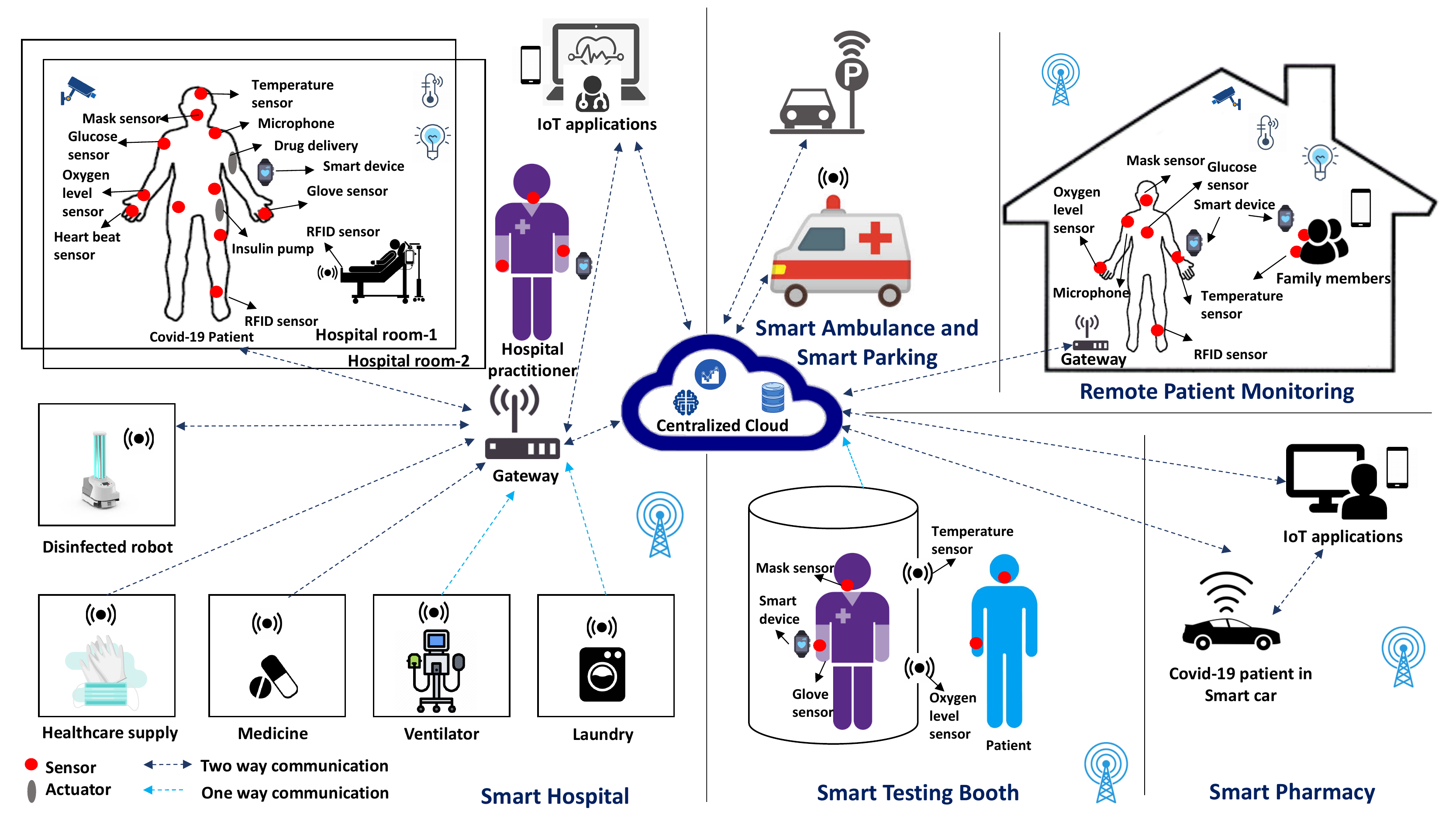}
\centering
\caption{Conceptual Overview of Connected Smart E-health Ecosystem.}
\label{fig:e-health}
\end{figure*}

It is expected that the global Internet of Medical Things (IoMT) market\footnote{https://www.forbes.com/sites/bernardmarr/2018/01/25/why-the-internet-of-medical-things-iomt-will-start-to-transform-healthcare-in-2018/\#57b39d4f4a3c} will grow to a 136.8 billion in year 2022. As of 2020, 3.7 million medical devices are in use today, which are used to monitoring patients’ conditions and sending data to hospital practitioners in real-time. E-Health helps to mitigate fast-spreading COVID-19 that has taken over the entire healthcare ecosystem including hospitals, testing booth, vaccine developing labs, pharmaceutical companies, and health insurers. Data-driven applications such as ML for drug discovery, RPM, predictive analytic for hospital resource optimization and interactive medicine deliver the right information at the right time in healthcare system. Improving the efficiency of healthcare services and keep healthcare costs under control have been an important and critical challenge during the COVID-19 pandemic time. To overcome these limitations, we discuss three important application scenarios: smart hospital, RPM, and smart testing booth, which are shown in Figure \ref{fig:e-health}.

%These use-cases involve connected smart sensors, connected devices, robots, patients, hospital practitioners, workers etc. together with IoT applications, edge devices and cloud services to offer data-driven services. Other scenarios including smart pharmacy, smart ambulance and smart parking in hospital's parking area, are also briefly discussed. Such scenarios can be extended with the current proposed architectures described in Section \ref{sec:model}.
\subsubsection{Smart Hospital} %Medical Facilities 
Today, smart hospitals are still facing several challenges including shortages of materials such as PPE and items that support a patient room, more specifically ventilator, thermometers, disinfectants, cleaning supplies, including hospital practitioners’ shortage in responding to the COVID-19 pandemic. To overcome these problems, we propose a smart hospital use case scenario which employs IoT technology to extend existing infrastructure to enable coordinated actions for coronavirus patients. Various components of smart hospital concept have also been studied in the literature~\cite{catarinucci2015iot,1213625,7070665,1504802}.
%Smart hospital \cite{catarinucci2015iot} uses RFID for
Within a smart hospital, RFID sensors can be an ideal way to keep track PPE, cleaning supplies, medical supplies, smart beds, ventilators, and patients. While RFID sensor is used to track the location of materials and patients within the hospital, other IoT sensors along with technologies including AI, big data analytics, and Blockchain can provide multiple benefits to patients and facilities. 

More specifically, hospital practitioners, staff members and patients can wear smart bracelets (e.g., iHealth smart watch) or smart rings (e.g., oura ring) to capture early symptoms of COVID-19 by collecting their data including body temperature, heart rate and blood oxygen levels and store at edge data centers, which avoids latency and security issues. 
This data can be used for training purpose to build models using Deep Learning (DL) or federated learning as presented in Algorithm 1 (as discussed in Section \ref{sec:implementation}). These models will increase the intelligence of data-assisted applications that can be used to predict early symptoms of coronavirus infection for hospital practitioners and staff members. Using IoT sensors and wearable devices, hospital practitioners can not only collect useful data with minimal in-person contact, but can also reduce the risk of cross-infection from the patients. Typically, due to limited storage on the edge, data can be sent to centralized cloud through secure data sharing tunnel. This tunnel is an encrypted link between edge device or gateway and centralized cloud. The searchable encryption scheme \cite{xu2019enabling} can be replaced with simple adoption of encryption scheme to enhance the privacy of IoMT data and also supports multi-clients.

The connected devices can communicate and operate autonomously. For example, smart ventilators can communicate through patient's smart bracelets with embedded sensors and can respond according to patient's body parameters. Health practitioners can also monitor patients and take decisions based on a combination of IoMT and Augmented-Reality (AR) technology. AR technology can provide additional benefits (e.g., breathing exercises) for patients. Autonomous robots can also help in healthcare systems as virtual clinic, smart guard and provide food service. Similarly, smart cameras can be employed for continuous monitoring of patients itself including hospital inventory, hospital resources, peak hours, etc. An autonomous mobile UV-C (ultraviolet-C) or UVD disinfection robot can disinfect rooms and equipment with ultraviolet light and Hydrogen Peroxide Vapor (HPV) and will traverse in hospital using random path planning algorithm to disinfect areas in swiftly. 

%As such having these IoT sensors and devices connected through edge computing and cloud platforms enabling an intelligent connected medical facility communication network of things as part of smart hospital are essential in such pandemic situations. 
%Besides, there are other IoT devices, such as disinfected robots which can autonomously disinfect a patient room regularly and after the patient is discharged, or a specific hospital area post contamination as needed. A study\footnote{https://spectrum.ieee.org/automaton/robotics/medical-robots/autonomous-robots-are-helping-kill-coronavirus-in-hospitals} discussed how UV Disinfection (UVD) robots are used to disinfect patient rooms and operating theaters in hospitals and trials are underway in Florida. 

\subsubsection{Remote Patient Monitoring (RPM)} RPM allows health practitioners to monitor COVID-19 patients remotely and provides care for recovering patients at home. Most popular companies, such as Gyant, Chronisense Medical, Ejenta, and iHealth, can enable RPM and provide a platform for physicians and hospitals to monitor patients outside of the hospital setting, where the technology can provide 24/7 data communication between patients and health practitioners. Patient-Generated Health Data (PGHD) including body temperature, SpO\textsubscript{2} percentage, BP level, pulse rate, coughing frequency, and ECG can be collected using wearable devices together with Internet-Enabled medical devices. This data will be sent to the web-server through Message Queuing Telemetry Transport (MQTT) protocol or Hypertext Transfer Protocol (HTTP) where it will be stored securely in patient’s private cloud or a specific medical cloud. PGHD can be retrieved for analysis by hospital practitioners enabling early detection of COVID-19 and identifying critical conditions of the patient using ML and DL algorithms. For example, if SpO\textsubscript{2} level of the patient is less than 85\%, hospital practitioners will receive alerts through data-driven application and then send extra care services to patient. Patient can also receive in-build corrective recommendations regarding medicines and extra precautions to be taken at home through IoT applications. He/she can also add his/her family members or friends to share his/her PGHD with different levels of access control due to security and privacy concerns. In general, due to privacy concerns, a patient might not agree to share his/her data, and in such cases, health devices will send data on a gateway device (e.g., patient's phone). The patient can also receive alerts based on threshold values through applications, especially when his/her health condition is critical. However, some alerts can be false positives, and to address such errors, dense sensor networks along with medical devices can help to reduce false positives. Furthermore, classification techniques such as Hidden Markov Model (HMM) \cite{deep2019survey} can be used to classify anomalies in PGHD. This model can be deployed on cloud or gateway to detect false alerts. 
%In addition, machine learning (ML) and deep learning approaches can be applied to this data to detect current and predicted critical conditions of a patient, and inform the hospital emergency staff members or autonomous smart ambulance regarding whether to bring the patient to the hospital. Family members can use IoT applications to monitor and communicate with the patient.

%Apart from the common wearable sensors, other smart devices including inhalers, insulin pen (only for diabetes patient), infrared body temperature, or connected indigestible sensors can be used in a E-Health scenario. The frequency of smart inhaler use by a patient can be monitored remotely through the cloud. If frequency exceeds a threshold, alert can be send to hospital practitioners. Similarly, family members of the patient can also be monitored by hospital practitioners to detect early symptoms of COVID-19. Local public health administration staff can only view the location of patient to monitor and enforce social distancing during self-quarantine. If patient violates the protocol of self-quarantine, smart applications can send alerts to local public health administration and hospital staff. Furthermore, if the person goes out during his quarantine time and breaks any guidelines, the  wearable device can initiate a notifications that sound repeatedly and loudly to inform the others in the vicinity. Local workers can access the records of patient’s history of movements to help them quickly identify the cross-infected areas and possibly infected coronavirus patients in a swift and timely manner.

\subsubsection{Smart Testing Booth} Boston hospital\footnote{https://whdh.com/news/boston-hospital-creates-coronavirus-testing-booth-designed-to-keep-healthcare-workers-safe-conserve-protective-equipment/} developed an innovative coronavirus testing booth to keep healthcare workers safe while swabbing possibly infected patients and conserve PPEs. A smart testing booth can include multiple sensors like infrared large-scale body temperature sensor, no contact oxygen level sensors with Red, Green, and Blue (RGB) camera, RFID scanners and AI assisted smart cameras. Single person at a time can enter one side of the glass-walled testing booth and will be identified through RFID tag based wearable device or optimized face recognition algorithm \cite{li2019optimization}. In-built sensors in the booth can record person's body temperature and oxygen level including other PGHD which will be stored in medical cloud. An individual can receive the result of COVID-19 test through a phone application and also set an alert on his/her wearable device for test results.

Some other IoT applications and services in the context of E-Health are discussed here. Smart ambulance can provide virtual on-board assistants to help patients and find optimal paths \cite{li2019special}, using ML services to the hospital, in case of emergency. The hospital smart parking system can scan license plate of driver to identify COVID-19 patient using patient's data and can also assign special spot to park his/her car. However, the patient's personal data can be stored on multiple storage to avoid centralization, which could lead to hacking, loss and mishandling of patient data due to centralization. Blockchain technology can be used to secure patient's personal data and can deliver decentralized attack detection model \cite{rathore2019blockseciotnet} to mitigate the single point of failure. 
%Another scenario depicts a smart pharmacy with non-contact curbside pickup. Generally, a pharmacist communicates with COVID-19 patient through an application and provides prescribed medicine to them. Before attending the other person, the pharmacist must take some time to sanitize the smart pickup box. [Not needed as already discussed many cases]
In the above scenarios, data and information collected from smart devices are sent to edge gateways, services, or cloud. Due to high security and privacy concerns in health domain, it is important to understand that these edge gateways and cloud-IoT platforms will be owned only by authorized entities, such as hospitals or other highly trusted entities through some private cloud. We elaborate some of the challenges to enable these scenarios in Section \ref{sec:challenges}.
\begin{figure*}[t]
\centering
\includegraphics[width=1.0\textwidth, height=.4\textheight]{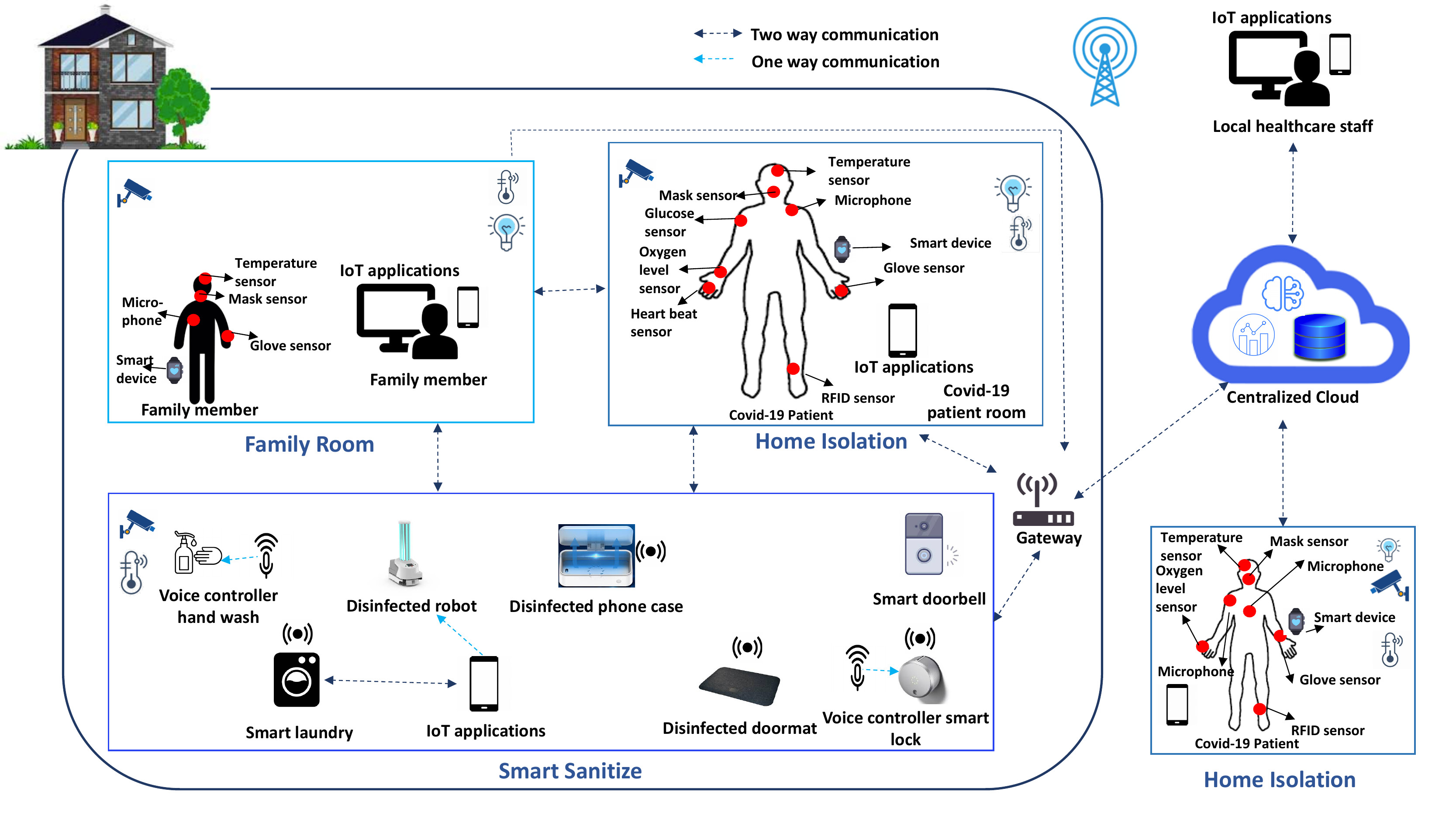}
\centering
\caption{Conceptual Overview of Connected Smart Home Environment.}
\label{fig:home}
\end{figure*}
\subsection{Smart Home} 
Today, smart homes make our lives easier and provide convenience in our daily activities from automatically turning on thermostat from smartphone to opening the smart garage door remotely. However, we can also utilize smart devices to keep our homes healthy and sanitized. In real-world, it is nearly impossible to keep everything around you virus or germs free. However, individuals must employ precaution measures to keep their homes clean and disinfected. In order to prevent the spread of COVID-19, the first step is to sanitize our homes. Currently, smart lights and temperature control with their compatible sensors have become ubiquitous in smart home environments enabled by Wireless Sensor Networks (WSN) \cite{bellavista2013convergence}. According to global tech market advisory firm\footnote{https://www.techrepublic.com/article/COVID-19-pandemic-impact-pushing-smart-home-voice-control-devices-to-predicted-30-growth/}, voice control device based orders and shipments will grow globally by close to 30\% over 2019. Using voice control allows people to avoid touching commonly touched surfaces around the home, such as smartphones, TV remotes, light switches, thermostats, door handles, and others. In this subsection, we elaborate three different scenarios within the context of smart homes which can offer a more connected ecosystem to mitigate contagious disease spread, as shown in Figure \ref{fig:home}. 
Moreover, various aspects of smart home for health have been investigated in the literature~\cite{7165043,6648859, 7501687,7574719,7206522,Okayode2020, gupta2020access,Tekeoglu2015,Tekeoglu2015b,Tekeoglu2016,Tekeoglu2017,Kayode2019,bhatt2017access, bhatt2017access1, bhatt2019authorizations}.

\subsubsection{Home Isolation} Some COVID-19 patients have high-risk enough to warrant quarantine but not serious enough to warrant smart hospital care. They should follow self-quarantine protocol and stay at their home; however, it will be a challenge to other family members to protect themselves in same home environment. In this scenario, it is expected that the patient should easily be monitored by hospital practitioners, family members and local state authorities with different level of authorization requirements. Patients' activities can be monitored through wearable and non-wearable devices at smart home. Non-wearable devices can be classified into three categories: vision-based devices (e.g., IR cameras, RGB cameras and depth cameras), medical devices (e.g., oximeter, scale, and thermometer) and dense sensing network based devices (e.g., door sensor, light sensor, and motion sensor). The patient and family members can use wearable devices (e.g., smart watch, smart mask, and microphone) to monitor their vitals regularly. Multiple sensors including mask sensor, temperature sensor, heart rate sensor etc. can be attached to patients body, and other smart health devices and smart home IoT devices can also be installed in the patient room, as shown in Figure \ref{fig:home}. PGHD will be collected from these smart devices, which are Bluetooth Low Energy (BLE) or Wi-Fi based devices, and this data will also be shared with health practitioners for data analytics. PGHD will be monitored by health practitioner through their private cloud or manufacturer's cloud (e.g., iHealth's cloud, home automation cloud). To enable trust across multiple clouds, Blockchain technology can be used. Additionally, Secure Data Transmission Scheme (S-DTS) \cite{kim2019dpn} can also support a distributed transmission among multiple clouds and can utilize four synchronization zones including IoT network zone, aggregation zone, mining zone, and management zone, and these zones can be operated by the Delegated Proof of Node (DPN) mechanism.

%At home, voice control assistant (e.g. Google home, Alexa) can help to patients 

%Patient and family members can control the environment (e.g., room temperature, lighting, humidity level for ease in breathing) in isolation and quarantine zone based on patients conditions, and at the same time members can also send reminders for medicines intake to the patient. For a home isolation situation, all family members are required to follow the guidance for implementing home care of the patient. In addition, these guidelines are updated frequently  and people can receive alert through voice assistants when new guidelines are available. 

\subsubsection{Smart Sanitize} 
Here, we present a smart sanitizing scenario which uses IoT technology. For instance, when a user will walk up to the home from the driveway and before he/she even reaches the front door, the door unlocks using optimize face recognition application \cite{li2019optimization} on the edge, then user will walk inside the home. User's phone will connect with home Wi-Fi network, and will show the presence of the user at home. The smart home will set to the right temperature, and turn on smart home devices without touching any surface. The user can take off his shoes on smart UV-floor, and put his keys, phone in smart UV light box that sanitizes these items. Hands-free automation including Gaggenau’s handle-free refrigerators, faucets, door and closets can also be used at home to avoid spread of virus. Voice-control devices such as Amazon echo, Honeywell Wi-Fi thermostat, hand sanitizer, and vocca light can be installed at home. 
Another futuristic technology that can be useful is air cleaning system that sanitizes clothing and footwear. This system can be installed in closets and laundry basket. Similarly, smart vehicles can be sterilized with steam. 
\begin{figure*}[t]
\centering
\includegraphics[width=1.0\textwidth, height=.4\textheight]{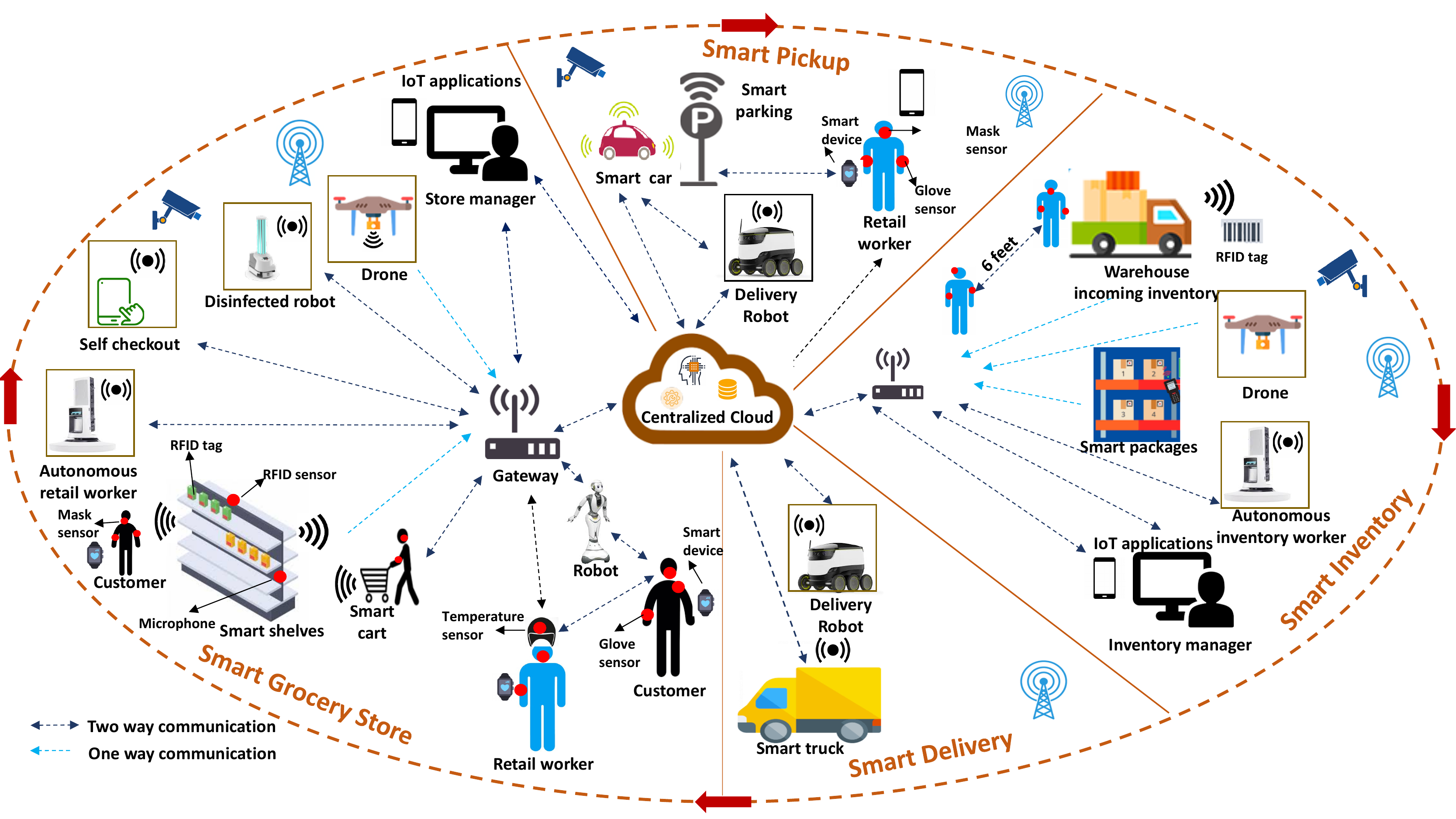}
\centering
\caption{Smart and Connected Supply Chain Management Scenario.}
\label{fig:supply}
\end{figure*}

\subsection{Smart Supply Chain Management} 
%During the pandemic, there is a requirement to improvise supply chain management to adapt automatic business processes and also improve the inventory with delivery of essential items.
COVID-19 pandemic has created many gaps in the supply chain management, the emerging IoT technologies can be deployed to improvise this whole system. The use case scenarios including smart inventory and smart grocery store are discussed in this section and also shown in Figure \ref{fig:supply}. These scenarios show how IoT devices and technologies can enable efficient supply-chain and also help in slowing the COVID-19 spread around us. 

\subsubsection{Smart Inventory} 
In the past research, smart inventory systems have been investigated in the literature~\cite{7750743,8780282,7917105,8939259,7932080,8666395,ding2013study}. The studies show that mostly RFID tags have been used to identify the shortage of goods in smart inventory. Inventories are facing an unprecedented challenge in coping with the fallout from COVID-19. However, a smart inventory system can provide a safe and secure environment to the workers using technologies. Within the inventories, drones can be used to track all the employees to check their temperature using thermal sensors, and also measure their social distancing using Algorithm 2 (as discussed in Section \ref{sec:implementation}). Inventory manager can also provide UWB/LTE-based wearable devices that are less susceptible to interference than Bluetooth devices, and workers' health generated data will be store on gateway server. If an employee is infected with coronavirus, then inventory manager can get notifications through data-driven IoT applications, given proper authorizations and privacy are enabled. In addition, disinfectant spray can be attached to the shelves that can start spraying when associated sensor senses the sound of sneezing. 
An intelligent drone can monitor inventory in real-time and send alerts in case there is a shortage. Stereo Vision, Monocular Vision, Ultrasonic, Infrared, Time-of-Flight and Lidar sensors are being used to detect and avoid obstacles. Sensor fusion can be used for obstacle detection, where data from different sensors are combined to compute useful information than could not be determined by just one sensor alone. Smart inventory will use RFID sensors and AI assisted drone to track the items, and RFID antenna/AI camera will scan the number of units on the sales floor and will send an alert to a store manager in case of shortage. The generated data will allow to automate product orders and also identify the popularity of a certain item. 

\subsubsection{Smart Grocery Store} 
Smart grocery store have been widely investigated in the literature~\cite{7932080,8743269,9040750,raad2018sysmart,1607945}. IoT along with AI technologies can help slow the spread of infection by enforcing prevention and detection mechanisms through connected sensors in a smart grocery store. %Due to the stay home order, people are panicking and stocking up grocery items. They need to stand in queues for hours outside the store to buy groceries. By employing IoT sensors around the store and wearable IoT devices, a store manager can get a better understanding to slow the spread in the store.  
Chinese tech firm Kuang-Chi Technologies\footnote{https://www.yicaiglobal.com/news/chinese-tech-firm-debuts-five-meter-fever-finding-smart-helmet} has developed a smart helmet attach with thermal sensor, which is used to take body temperature at the retail store. 
Similarly, thermal cameras\footnote{https://spectrum.ieee.org/news-from-around-ieee/the-institute/ieee-member-news/thermal-cameras-are-being-outfitted-to-detect-fever-and-conduct-contact-tracing-for-covid19} and microphone sensors can also be installed at the store which can detect people who are coughing in store during shopping. %These areas will be disinfected and identified individuals may be reported for testing based on their other symptoms and will be categorized as risky customers. This information can be maintained in store for short-period of time to assist in identifying these individuals during their future visits to the store. 
The customer can wear a RFID tag or smart bracelet at the store, which can be provided by state government during COVID-19 testing. The customer can scan the tag/bracelet at the entrance of grocery store, which will allow him/her to enter based on his/her health condition. From a customer's perspective, the user can enable alerts on his smartphone regarding his grocery list, can see the map of the store and crowded aisles, and plan accordingly to maintain social distance while shopping. The customer can visit desired aisles and will get items from the smart shelves that will put items in the smart cart without contact. Smart shelves will have three common elements - an RFID tag, an RFID reader, and an antenna. The customer generated data will be collected by smart shelves and smart cart during the day and this data will be analyzed on retail cloud. Social distancing can be measured and enforced by AI assisted autonomous retail robots as well, who can alert the customers through speaker. The UVD Robot can also use ultraviolet light to zap infection viruses and sanitize surfaces.
%and shopper buying trends, patterns, shopper traffic, etc. will be shared with a store manager to provide customer-related insights to efficiently manage the store inventory and restocking goods. Most retail stores now allow only ten people at 30 minutes shopping interval slot to avoid large gatherings inside the store. If two customers come in same aisle and violate the physical distancing norm, an autonomous retail worker will go there to warn them or a loudspeaker attached with smart camera will announce to keep maintain social distancing. Autonomous retail worker can roam around the store and can take note of misplaced items, or products running out of stock (smart shelves also keep track of items and can send alert for restocking as needed). AT\&T\footnote{https://www.fiercewireless.com/iot/at-t-4g-lte-connects-iot-robots-to-kill-germs-keep-shelves-stocked} with Xenex and Brain Corporation has already launched autonomous robots to help grocery stores in keeping them clean, killing germs and maintaining well stocked shelves more efficiently. 

Now-a-days, most grocery stores provide contact-less drive through pickup for all customers and delivery at their doors. However, some users, such as elderly people, may not be aware or familiar with such online options. For them, grocery store can provide smart robot delivery or smart pickup outside the store and also provide a separate parking area to avoid long waiting lines. %Smart pickup can automatically allow most vulnerable and infected people first and enforce these rules to inform the vehicles. 
A restaurant takeout service can follow the same protocol for smart pickup. Intelligent Transportation System (ITS) \cite{gupta2020secure, 9187899, gupta2019secure} can support in delivering resources to users. Gupta et al. \cite{gupta2020enabling} have also elaborated how ITS and smart city infrastructures can be used to enable and enforce social distancing community measures in COVID-19 outbreak.
\begin{figure*}[t]
\centering
\includegraphics[width=1.0\textwidth, height=.4\textheight]{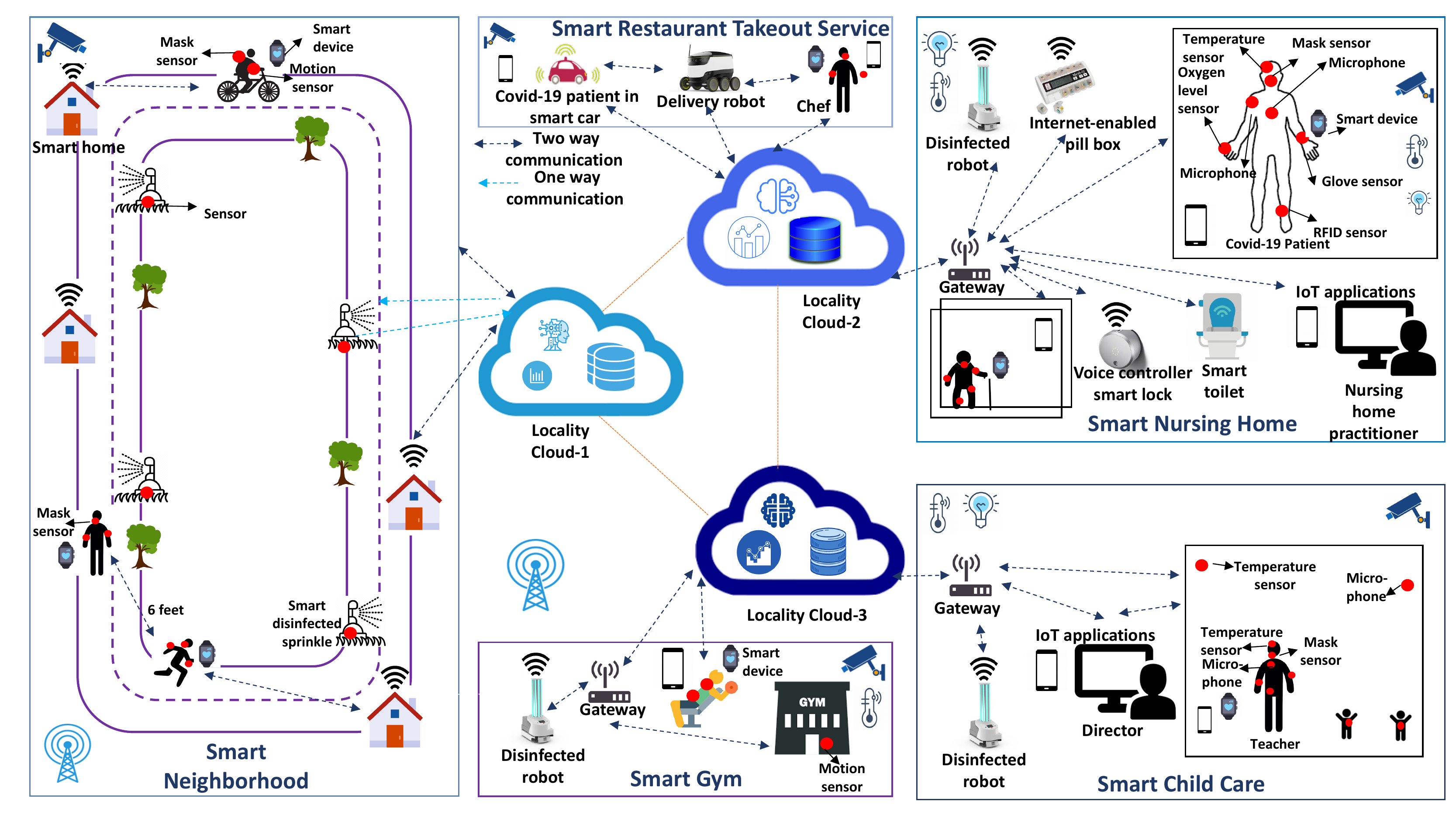}
\centering
\caption{Connected Future Smart Locality.}
\label{fig:community}
\end{figure*}
\subsection{Smart Locality} Smart localities have been widely investigated in the literature~\cite{6069711,7406686,8355907,7448556,8656893,7384112}.
Such localities consist of various interdependent human and physical systems, where IoT represents the sensing and actuating infrastructure to estimate the state of human and physical systems and also assist in adapting/changing these systems. Here, we discuss two scenarios, which can help humans to adapt to the new normal in COVID-19 situation. These two scenarios including other relevant scenarios are shown in Figure \ref{fig:community}.

\subsubsection{Smart Neighborhood} Every individual, who lives in a smart neighborhood, will receive notifications regarding allotted time for outside activities, such as riding a bike, a walk on the trail, etc. in order to maintain the social distancing while being outside in common areas. In a smart neighborhood, motion sensors and cameras will sense the presence of people and will send the counting numbers to the locality cloud-1 as shown in Figure \ref{fig:community}. Cloud Smart Analytics\footnote{https://cloud.google.com/solutions/smart-analytics/} service can analyze the locality data and will send notifications to users. IoT application can suggest different paths for each individual using motion sensors and traversing algorithms, so that people can avoid meeting infected people without knowing his/her identity. 
%to people regarding number of positives cases and categorize risk zones with different colors (e.g., red - high risk, yellow - medium risk, and green - low risk) in smart neighborhood. When a person will go for a walk in a smart neighborhood, he/she will receive an alert if any infected person/pet are around and also alert them to avoid high risk (red) zones in smart neighborhood. 
There will be disinfected sprinkler installed that can spray on the pedestrian path and community park, when sensors will sense the presence of infected person in the area through notifications from the locality cloud-1. 

\subsubsection{Smart Child Care and Nursing Home} The U.S. CDC\footnote{https://www.cdc.gov/coronavirus/2019-ncov/community/schools-childcare/guidance-for-childcare.html} have issued guidelines in order to safely open child care and schools in the locality. The child care can use RFID tags, motion sensors, door sensors, and smart devices for children safety. The sensors can be attached to kid's clothes to monitor their sleep patterns, breathing patterns, body temperature, and body position and this data can only be stored on edge server for child security purpose, while staff/teacher generated data can store on locality cloud-3. AI assisted camera can be installed in child care rooms to measure social distancing among children and teachers and it is shown in Algorithm 2. However, it is a challenging undertaking to maintain physical distancing in child care. In such scenario, sensor based partitions of each room will allow a gap among the kids and will allow only one or two kids in each partitioned area to maintain the distance among them. The teacher and staffs will receive notifications to change their gloves and masks through attached sensors on gloves and mask. In a smart nursing home, if an elderly person shows symptoms of COVID-19, then he/she should be treated in an isolated room immediately. Autonomous robots can be utilized to help in disinfecting common areas, and by following Algorithm 3 (as discussed in Section \ref{sec:implementation}). Elderly health generated data and their clinical reports can be accessed through applications and will store in locality cloud-2. The care coordinator should have access to locality cloud-2 to monitor each elder regularly. 
%In a smart child care, it should be mandatory to wear mask and gloves for all teachers and staff as well as for kids throughout the day. They can receive a notification to change their mask and gloves through IoT application using mask sensor and glove sensor, when their masks and gloves are infected from virus and germs or after certain time period. The smart child care director and administrators can monitor body temperature, frequency of coughing, oxygen level of teachers, staff members, and kids through the application and get insights from ML techniques. 

%Coronavirus-linked fatalities at nursing homes and other long-term care facilities in the U.S. have surpassed 10,000 in New York, according to news report\footnote{https://nypost.com/2020/04/23/coronavirus-deaths-at-us-nursing-homes-reach-over-10000/}. Japan’s artificial intelligence expertise\footnote{https://thediplomat.com/2018/06/japans-robot-revolution-in-senior-care/} is transforming the elder care industry, with niche robotic caregiving accomplishing more than just taking pressure off the critical shortage of caregivers.
 
%other people and staff members,  Figure \ref{fig:community} shows that the body parameters of the elderly people can be taken through attached body bracelet and sent to edge devices and gateways. IoT devices and applications connected through locality cloud-2 can share information of COVID-19 patients to smart hospitals. The nursing home staff can monitor the  patient's body parameters regularly, and will also track other elderly people in the smart nursing home. 
Similarly, a smart gym would include motion sensors, door sensors, humidity sensor, temperature control sensors, and AI assisted cameras. The generated data will be stored on locality cloud-3, and will be analyzed by cloud services. Gym manager/HOA can only access the analyzed information anonymously, so that he can assign a particular time to maintain the 25\% occupancy and time interval to sanitize all gym equipment and surfaces. To enable multi-cloud secure data and information sharing and communications between locality clouds, there is a need to build decentralized trust framework in place of using advanced technologies like Blockchain and trusted distributed computing.
%A gym member will receive a notification regarding to come to the gym, it must require maintaining the 25\% occupancy at the gym and time interval to sanitize all gym equipment and surface.   

\subsection{Smart City}

\begin{figure*}[t]
\centering
\includegraphics[width=1.0\textwidth, height=.4\textheight]{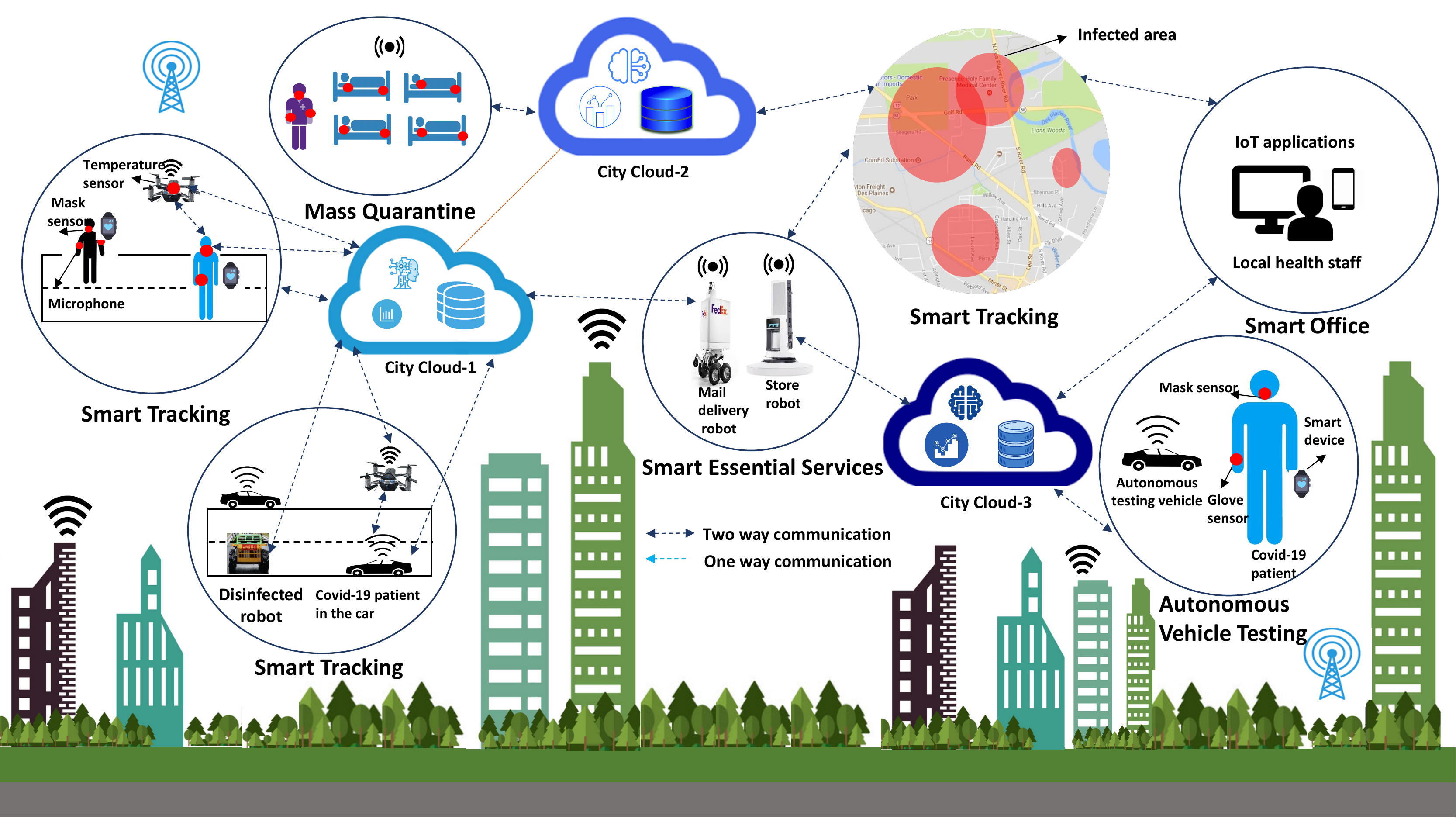}
\centering
\caption{Conceptual Futuristic Smart City Overview.}
\label{fig:city}
\end{figure*}

Daegu\footnote{https://www.smartcitiesworld.net/news/news/south-korea-to-step-up-online-coronavirus-tracking-5109} has setup a novel system using large amount of data gathered from various sensors and devices, such as surveillance camera footage and credit card transactions of confirmed coronavirus patients to recreate their movements. The Newcastle University Urban Observatory\footnote{https://COVID.view.urbanobservatory.ac.uk/\#intro} developed a way of tracking of pedestrian, car parks, traffic movement to understand how social distancing is being followed in Tyne and Wear. However, other major cities need to prepare themselves for coronavirus future outbreak waves. 
%Voxel51\footnote{https://pdi.voxel51.com/} is tracking the impact of the coronavirus global pandemic on social behavior, using a metric they developed called the Voxel51 Physical Distancing Index (PDI). Using the cutting-edge computer vision models and live video streams from some of the most visited streets in the world, the PDI captures the average amount of human activity and social distancing behaviors in major cities over time.
%[ their health policies and quarantine protocols are observed to diverge from one another.] 
Smart city can host a rich array of technological products that can assist in early detection of coronavirus outbreaks through IoT sensors, and mitigate infection through social distancing and mandatory masks policy. Connected sensors and data sharing through the cloud-IoT architectures can be utilized to enforce social distancing measures around a smart city. Here, we discuss various scenarios including smart tracking, and autonomous testing vehicles for the smart city, which are shown in Figure \ref{fig:city}.
In the past, several research work \cite{mora2018use,6740844,6702523,7721743,6871673,7406686,7876852,9084093,9018282} have been done on smart city infrastructures.

\subsubsection{Smart Tracking} 
Most of countries\footnote{https://www.businessinsider.com/countries-tracking-citizens-phones-coronavirus-2020-3} used phone data to track citizens’ movements during the pandemic. In addition, urban cities deployed connected drones and law enforcement, which can help to track and restrict movement of individuals in a smart city without evading citizens' privacy. A smart city can be characterized by smart infrastructures that will be optimized for performance to mitigate coronavirus outbreaks and saving lives of the citizens. 
%The whole idea of a futuristic city when it comes to deployment of architecture having 
The smart infrastructures need to follow the protocols defined by federal and state agencies. In Spain, Nice city\footnote{https://www.businessinsider.com/coronavirus-drones-france-COVID-19-epidemic-pandemic-outbreak-virus-containment-2020-3} have been using drones controlled by police to enforce the lock down and circulate information announced messages like \textit{"Please respect the safety distances."} attached with loudspeakers. These Unmanned Aerial Vehicles (UAVs) also sprayed disinfectant at Zhengwan village, amid an outbreak of the new coronavirus, in Handan, Hebei province, China. Savannah city\footnote{https://thehill.com/changing-america/resilience/smart-cities/492532-savannah-ga-to-enforce-social-distancing-with-drones} announced plans to use drones to enforce social distancing among residents. 

In smart city, AI assisted drones are being used to measure six feet distance among people using distance measurement sensor mono/stereo. The distance measurement sensors are divided into multiple categories; however mono/stereo are less costly, accurate, fast, and high resolution. AI assisted drones can also identify those people who do not follow the protocols, for example, user is not wearing a mask, may receive \$500 fine. In smart cities, multiple drones monitor citizens, to avoid collision obstacle avoidance algorithm and Simultaneous Localization And Mapping (SLAM) can be employed. These drones train their models on edge using federated learning process before sharing its training data with an edge or city cloud-1, federated learning algorithm is discussed in Section \ref{sec:implementation}. Thus, each drone can monitor its area based on individual experiences. However, aggregate value of parameters from nearby urban area is collected at the edge computing side for collective decision making. 

%Thermal sensors can also be attached to the drone to check the body temperature of people.

As shown in the Figure \ref{fig:city}, COVID-19 test result of a patient can be stored on city cloud-1, which can be shared with local state staff and healthcare providers. Most dominant Cloud Service Providers (CSPs) use their own default data placement strategy to manage storage and computing of data, the CORE-Algorithm \cite{vengadeswaran2019core} can be use to re-organizes the distributed file system by redistributing the data blocks to give an optimal data placement. Furthermore, QR code can help the citizens using the geographic data to know about risk zones in the city.

%Other smart devices including drones, traffic lights and road side sensors based on a need basis. This will enable patients to be easily identified using face recognition while they are in visiting different locations in a city. 

\subsubsection{Autonomous Testing Vehicle} 

In order to keep patients and healthcare providers safe, drive-thru coronavirus testing sites have been popping up in the city. An autonomous testing vehicle can be used for COVID-19 testing in urban city as well as rural areas. Multiple sensors including infrared body temperature sensor, oxygen level sensor, AI assisted camera and microphone will be installed onto the vehicle which will also carry testing kits, PPE etc. This autonomous vehicle can help reduce exposure of elderly and pre-existing condition people to COVID-19. A person will get notification of autonomous vehicle's arrival, then he/she can enter from one side of the glass-walled area in the car. In-built sensors can record person's body temperature, oxygen level and can store this data on local edge device. The generated health data will be stored on local edge storage and these sensors will also publish the data on city cloud-3 to share with healthcare practitioners. It can also provide a test kit to individuals who can test themselves and return it through the car window. In rural areas, autonomous testing vehicles can bring awareness of COVID-19, improve accessibility to testing, and can even provide free testing to underprivileged communities. 

To flatten the curve of confirmed cases, smart city can provide mass quarantine for coronavirus patients, who have mild symptoms but are higher risk individuals for spreading cross-infection to others. A smart hall or large stadium or facilities can be setup for quarantine with installed sensors, smart devices, and robots that are connected to cloud (as shown in Figure \ref{fig:city}). Disinfectant robot is an autonomous robot that can sterilize floors in these large areas as discussed in other scenarios. The large-scale disinfectant robot can also be used to clean the roads of the city. Autonomous and self-driving vehicles can be used for delivering the post, which will also help reducing the human contact and cutting down the number of COVID-19 cases. Smart city can also provide immunity-based RFID tags to those people, who recover from the disease, and can allow the tag holders to return to work with extra-precautions. In the future, once COVID-19 vaccines are available, the individual with vaccination can get similar immunity-based RFID tags to prove their immunity. In other context, autonomous vehicles have been extensively investigated in
the literature~\cite{554205,250509,1306972,791202,4162483,7585053,7139555,gupta2019dynamic,gupta2020dynamic}. 
\begin{figure*}[t]
\centering
\includegraphics[width=1.0\textwidth, height=.38\textheight]{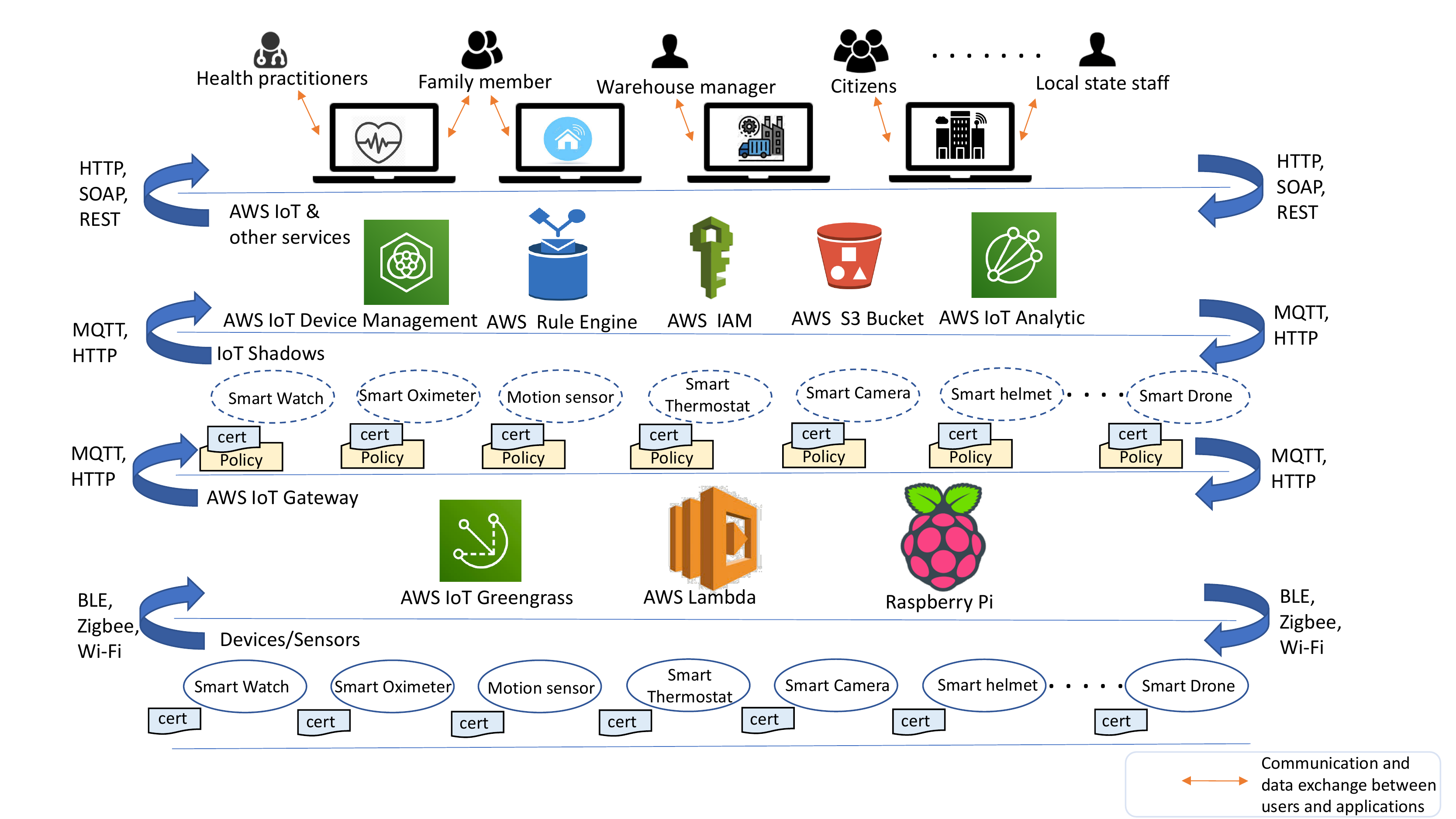}
\centering
\caption{General Implementation Framework for Smart Communities utilizing AWS IoT and Cloud Services}
\label{fig:AWS}
\end{figure*}

\section{Implementation and Enforcement}
\label{sec:implementation}
In this section, we present a general implementation and enforcement framework for smart community use case scenarios discussed in Section \ref{sec:usecases} utilizing AWS cloud and IoT platform. We also discuss a family of algorithms that can be used in the context of different use case scenarios. To further demonstrate the applicability and benefits of smart connected communities, we also implement a RPM use case along with proof of concept deployment setup, configurations for physical devices, and cloud services used for enforcing this scenario. This prototype implementation can be considered as a baseline which can be extended and customized to wide geographic and large scale connected communities. It must be noted that AWS is one of the cloud providers that can support this general framework. Similarly, other architectures can also be supported using different cloud and IoT providers, such as Google IoT\footnote{https://cloud.google.com/iot-core}, Microsoft Azure\footnote{https://azure.microsoft.com/en-us/}, Openstack\footnote{https://www.openstack.org/} etc.

\subsection{Generalized Implementation Framework in AWS}
The smart connected ecosystem with specific use case scenarios can be implemented using real-world cloud-enabled IoT architecture, such as supported by AWS cloud and IoT platform. Figure \ref{fig:AWS} illustrates our proposed implementation architecture, which can be used as a general framework for the use case scenarios discussed in Section \ref{sec:usecases}. This framework consists of smart devices per use case requirements, virtual things/objects, AWS cloud and its IoT and other services. Figure \ref{fig:AWS} shows a set of devices at object layer and their corresponding virtual objects at VO layer (as discussed in multi-layered architecture in Section \ref{sec:model}). However, other types of devices and their VOs can be created based on use case scenarios. More specifically, this proposed implementation includes AWS IoT core\footnote{https://docs.aws.amazon.com/iot/latest/developerguide/what-is-aws-iot.html} in cloud, AWS Greengrass\footnote{https://aws.amazon.com/greengrass/} running on edge gateways (e.g., Raspberry Pi enabled gateway), and IoT devices/things. The edge gateway can be enabled using Raspberry Pi\footnote{https://www.raspberrypi.org/} that host the AWS Greengrass deployment, and customized lambda\footnote{https://aws.amazon.com/lambda/} functions that can run on the gateway for edge communication and computation, sending notifications, or enforcing access control and privacy policies. After setting up AWS Greengrass gateway, data collection from smart devices can be processed locally to enhance the real-time intelligence of IoT applications and devices, critical in health related scenarios, such as RPM. For instance, a COVID-19 patient's body temperature and oximeter data need to be monitored regularly, and if there are any complexities (e.g., high temperature, breathing issue), then a notification would be sent to a connected ambulance. The patient can also define more restrictive policies for data security and privacy, for example, only hospital doctor can access his/her health data. 

AWS Greengrass can enable edge communication between devices and gateways. As shown in Figure \ref{fig:AWS}, AWS Greengrass allows users to securely connect their devices to gather data and communicate with gateway (where Greengrass is deployed) which allows to take intelligent actions locally even when the internet connectivity is down. When connectivity is re-established, the data will synchronize with other cloud based IoT services including AWS IoT core, AWS IoT device management, AWS S3\footnote{https://aws.amazon.com/s3/} bucket etc. Data collected by smart devices can also be forwarded to AWS IoT cloud using MQTT/HTTP protocols. For enabling a secure architecture, AWS IoT core provides mutual authentication and encryption for secure data transfer and at rest. The mutual authentication mechanism is based on X.509 certificate to ensure proper secure device provision and authorization for data exchange between devices and AWS IoT core. AWS IoT will generate certificates signed by certificate authority, which can be assigned authorization policies to secure user devices and data, and also to enable different access levels (full control/view only) to individuals (e.g., family members, friends, and health practitioners). In AWS IoT, device management service allows users to quickly on board their large and diverse fleets of devices and store the data while keeping their fleets secure. Some other services on top of them like AWS IoT analytic can also be used to analyze large amount of data collected from these devices. For instance, health practitioners can analyze patient's health data using AWS IoT analytic service. It can also integrate seamlessly with other AWS services, such as amazon quick site for visualization of IoT data and AWS sage maker hosted by ML service which is used for predictive analytics. In this proposed framework, the smart devices (e.g., smart watch, oximeter etc.) must be compatible with AWS cloud to use its services. 
%Various scenarios that can be implemented using this architecture, where data associated with body parameters of people is sent to edge gateway device, which is then forwarded to the virtual objects (i.e., Device Shadows in AWS) in AWS IoT platform. Their health data will be stored in AWS S3 bucket which will be continuously analyzed and monitored by health practitioners. If provider realizes any early symptoms of COVID-19,  a lambda function or a rule can be triggered to initiate appropriate action based on the data values. 
%For example, staff members will take early precautions and will assign an isolation room to that elderly person.

\begin{spacing}{1}
\begin{algorithm}
\label{alg:fed}
%\SetAlgoLined
\caption{Federated Learning on Smart Edge Gateway Device i for Monitoring}
\begin{algorithmic}[1]
\STATE{Define initial parameters $w^i$, learning rate $\alpha$ and number of local epochs H.}
\STATE{Repeat all the steps until minimum error is obtained:}
\STATE{Download parameters $w^{global}$ from Parameter Server to learn the symptoms of COVID-19.}
\STATE{Collect user health data from wearable devices including smart bracelets, smart rings and non-wearable devices including thermal scanner, oximeter and send the data to edge devices including mobile phone, Raspberry Pi.}
\STATE{LocalTraining (i,$w^i$): Train user health data on each device to build own ML model for monitoring.}
\STATE{Split local dataset $D_i$ to minibatches of size K which are included into the set $K_i$.}
\FOR{each local epoch j from 1 to H}
    \FOR {each k $\in$ $K_i$}
\STATE\begin{equation}
\Delta w^{i} = \Delta w^{i}  - \alpha\frac{\partial E_{i}}{\partial w^{i}}
\end{equation}
\ENDFOR
\ENDFOR
\STATE{Each edge device uploads local gradients $\Delta$$w^i$ to Parameter Server.}
\STATE{Parameter Server also aggregates these local gradients $\Delta{w^i}$ asynchronously.}
        \begin{equation}
        w\textsuperscript{global} = w\textsuperscript{global} + \Delta w^{i}
        \end{equation}
\STATE{Each edge device downloads updated parameters $w^{global}$.}
\end{algorithmic}   
\end{algorithm}
\end{spacing}

We have proposed three algorithms which can be adapted and aligned based on the needs of different smart community use-cases. Algorithm 1 outlines a wearable and non-wearable device’s federated learning process on the edge including mobile phone and Raspberry Pi before sharing its training data with cloud. We assume local vector of neural-network parameters $w^i$ to learn a model to predict the symptoms of COVID-19. Parameter Server (PS) maintains a separate parameter vector $w^{global}$. The model on edge device will initialize parameters (weights) $w^i$, where \textit{i}=1,2,3,..\textit{N}, randomly or by downloading their latest values from PS. Each edge device or Raspberry pi trains a local model and optimize the loss value using Stochastic Gradient Descent (SGD). This process continues until SGD converges to a local optimum. Let E be the error or loss value, i.e., the difference between the true value of the objective function and the computed output of the network, and it can be based on $L^2$ norm or cross entropy. The back-propagation algorithm computes the partial derivative of E with respect to each parameter in $w^k$ and updates the parameter so as to reduce its gradient. We refer to one full iteration over all available input data as an epoch. Thus, the mobile device and Raspberry Pi train user's data on edge. Algorithm 2 outlines that smart assistant camera, attached with distance measure sensors including mono/stereo, measure the distance $\delta$ between two users at any smart community to enforce social distancing. The user will receive alert, if his distance from other user is less than a certain threshold value (e.g., 6 feet). 

\begin{spacing}{1}
\begin{algorithm}[H]
\label{algo:social}
%\SetAlgoLined

\caption{Social Distancing through Smart Assistant Camera i}
\begin{algorithmic}[1]

\FOR{each smart assistant camera i}
\STATE{Measure $\delta$ using mono/stereo distance sensors attached in smart assistant camera}
    \IF {$\delta$ $<$ predefined threshold}
    \STATE{Send alert to maintain distance}
    \ENDIF 
\ENDFOR 
 \end{algorithmic}  
\end{algorithm}
\end{spacing}

\begin{spacing}{1}
\begin{algorithm}[H]
\label{algo:traverse}
%\SetAlgoLined
\caption{Traversing Strategy for Smart Sanitizing Device i}
\begin{algorithmic}[1]
\STATE{Divide any infrastructure into three different zones: Isolation Zone (IZ), Living Zone (LZ), and Staff Zone (SZ)}
\STATE{Call count\_person()}
\FOR{each smart sanitizing device i}
\IF{number\_IZ $\ge$ max-threshold-1}
\STATE{sanitize IZ}
\ENDIF
\STATE {$\alpha$ = max(number\_LZ, number\_SZ)}
\IF{$\alpha$ $\ge$ max-threshold-2}
\STATE{sanitize that zone}
\ENDIF
\ENDFOR 
\end{algorithmic} 

\begin{algorithmic}[1]
\STATE\textbf{count\_person()}
\STATE{set number\_IZ= 0, number\_LZ = 0, number\_SZ = 0}
\FOR{each motion sensor j}
\IF {mono/stereo, ultrasound and motion sensors detect presence of person in IZ at any infrastructure}
\STATE{number\_IZ = number\_IZ + 1}
    \ELSE
        \IF {mono/stereo, ultrasound and motion sensors calculate number of person at LZ at any infrastructure}
            \STATE{number\_LZ = number\_LZ + 1}
        \ENDIF
    \ELSE
        \IF {mono/stereo, ultrasound and motion sensors detect the persons in SZ at any infrastructure}
        \STATE{number\_SZ = number\_SZ + 1}
        \ENDIF
\ENDIF 
\ENDFOR
\end{algorithmic} 
\end{algorithm}
\end{spacing}
Similarly, Algorithm 3 presents a general traversing strategy that can be followed by any smart sanitizing device within any smart infrastructure including smart hospital, smart nursing home, and smart home etc. We assume that smart infrastructure can be divided into three different zones: Isolation Zone (IZ), Living Zone (LZ), and Staff Zone (SZ), where COVID-19 patients live in IZ, the presence of visitors and working staffs represent by LZ and SZ respectively. The threshold value of IZ will be different form LZ and SZ. The motion sensors will count the number of person at a particular zone. The max-threshold-1 is assigned for IZ, while max-threshold-2 is assigned for LZ and SZ. Smart sanitizing devices including autonomous robots will communicate with motion sensors and will traverse accordingly as per the algorithm.
Here, we present generic algorithms that can be adopted in different scenarios as well as can be customized as per the needs of specific application scenarios. Discussed use case scenarios can implement any of the defined algorithms on AWS cloud based implementation and enforcement architecture with other technologies. 
%The certificate is attached with each device for authentication.

%The generated data from smart devices can send to AWS IoT greengrass which allows to execute AWS lambda functions locally at the edge. For instance, if the patient has high temperature 102\si{\degree} F, then a customized lambda function can trigger thermostat to desired temperature, such as 80\si{\degree} F). 

%AWS IoT greengrass synchronizes the required files to the Raspberry Pi. Thermostat runs as a trigger, when a patient has high temperature 102\si{\degree} F and feeling cold, smart thermometer sends a message to AWS IoT Core. AWS IoT Core invokes the recognition Lambda function, which is deployed on Raspberry Pi local storage, and if the Lambda function triggers thermostat and sets at 80\si{\degree} F. -- please rephrase this part if you can, currently it does not make sense, otherwise we need to remove it.

%Another scenario showing the data flow in the use case is discussed here. A smart oximeter measures Bob's blood oxygen level, pulse rate, and sends data to AWS greengrass, which is then sent to its thing shadow in AWS IoT platform. Bob's oxygen level data is stored in AWS S3 bucket and is analyzed and monitored by health practitioners continuously. If health practitioners realize that there is an emergency or a lambda function or rule is triggered based on the data values, then an ambulance can be sent to his home. Family members who are present at home, they can also monitor Bob's health through remote patient monitoring application.
\begin{figure*}[t]
\centering
\includegraphics[width=1.0\textwidth, height=.38\textheight]{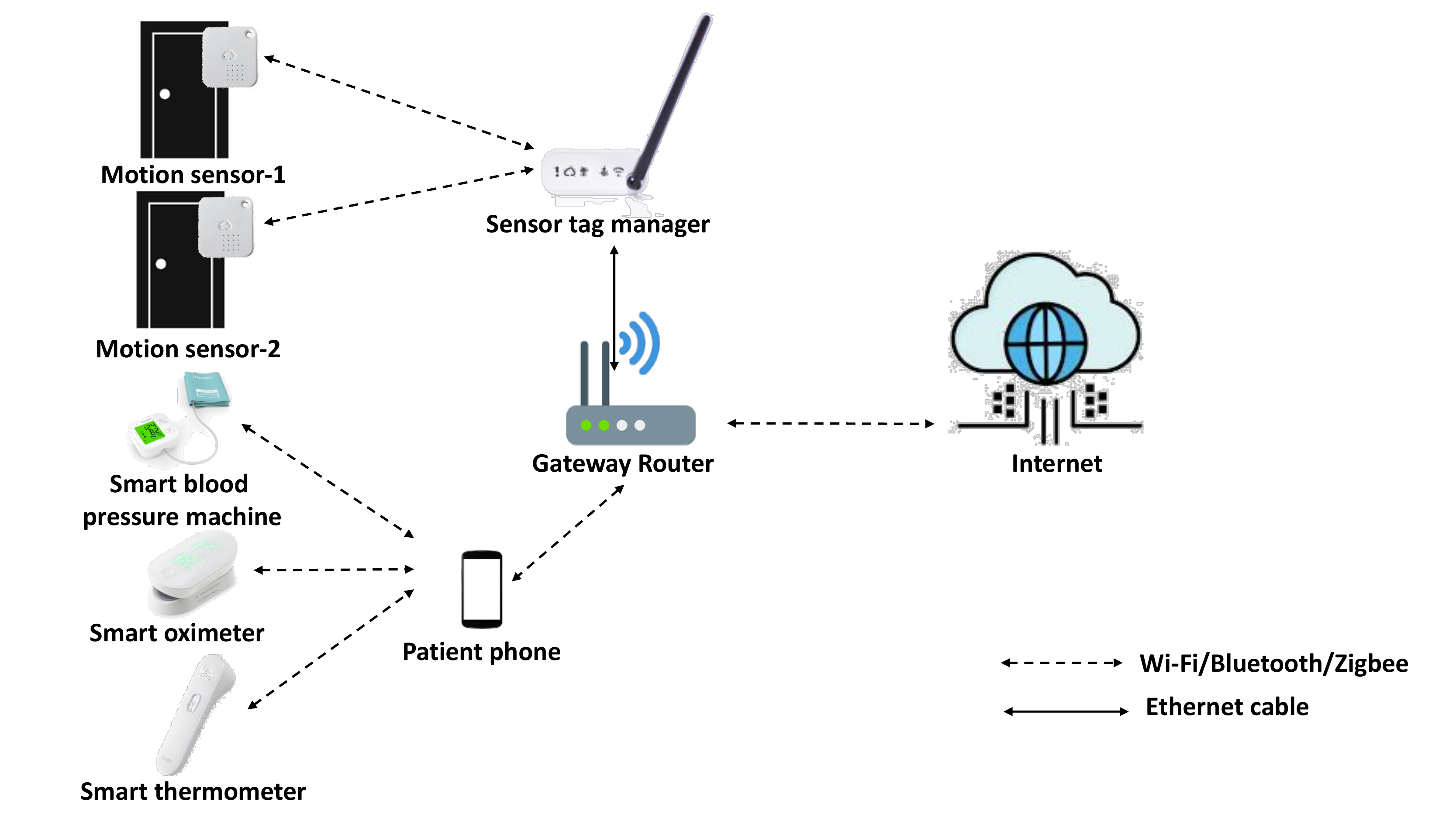}
\centering
\caption{Set up of Remote Patient Monitoring in a Smart Home}
\label{fig:setup}
\end{figure*}

\begin{figure*}[t]
\centering
\includegraphics[width=1.0\textwidth, height=.38\textheight]{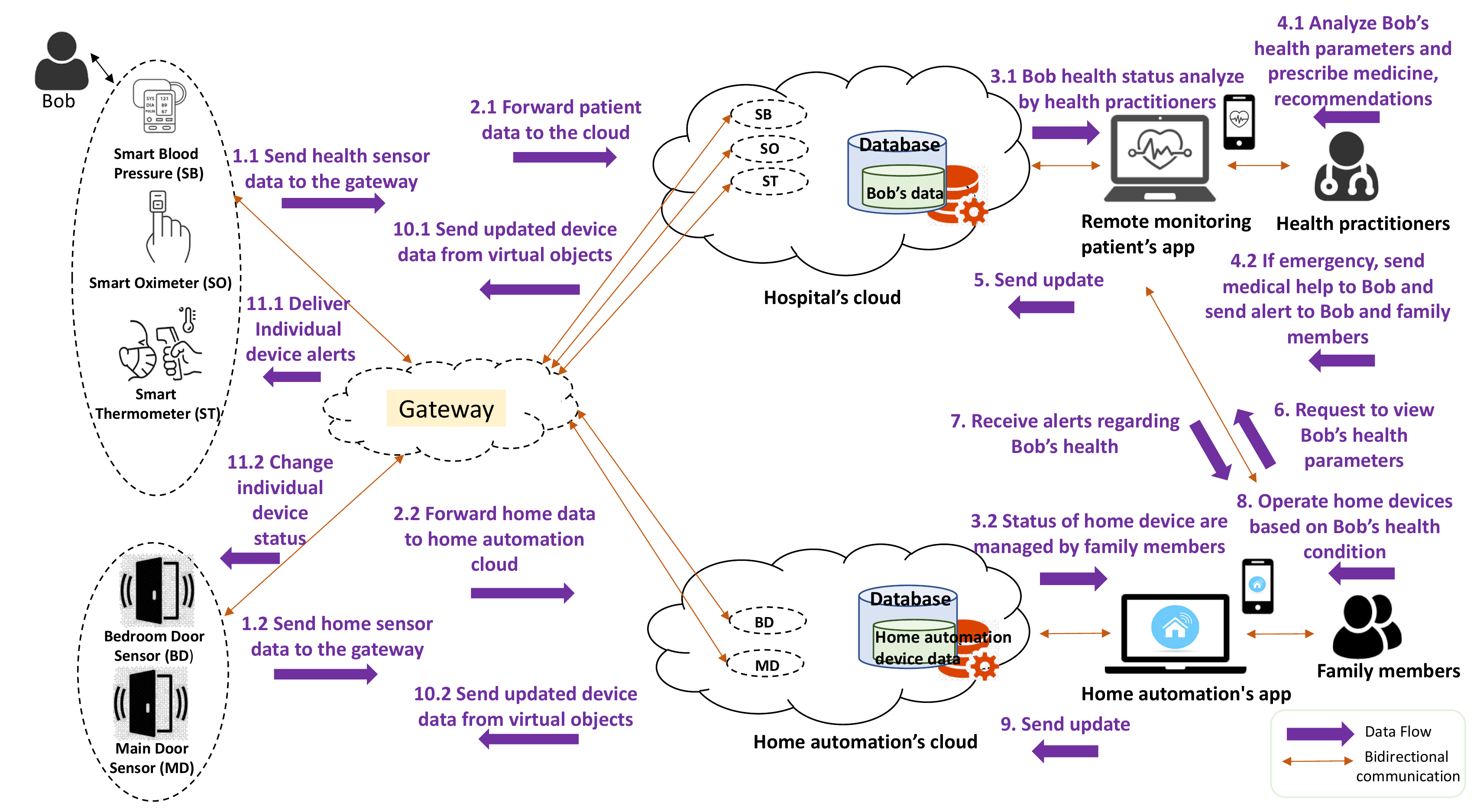}
\centering
\caption{Sequential View of Remote Patient Monitoring in a Smart Home}
\label{fig:RMP}
\end{figure*}

\subsection{Proof of Concept Implementation of Remote Patient Monitoring Use Case}
E-Health use case consists of various scenarios including smart hospital, RPM, and smart testing. This section presents implementation of a RPM scenario in real-world. We consider Bob, who is 34 years old and lives alone at his home, has been diagnosed with coronavirus. He is monitored by his health practitioners continuously through RPM system. His family members, who live in a different state, also monitor him through RPM system. We set up and configure the RPM system with smart health and smart home devices at his home setting as shown in Figure \ref{fig:setup}. For implementation of RPM, we set up a wireless network at Bob's home using ARRIS SB6121 modem, TP-Link AC1200 wireless router, and smart devices including Ethernet Tag Manager, wireless sensor tags, and various BLE based health devices (e.g., iHealth smart oximeter, iHealth smart blood pressure monitor, and iHealth smart thermometer). These health devices are suggested by Bob's health practitioners. Ethernet tag manager is connected to router through CAT 5 Ethernet cable, it is linked with two wireless sensor tags installed at bedroom door and main door. This entire test bed set up is deployed around 8-10 days.

Figure \ref{fig:RMP} shows a sequential view of RPM for COVID-19 patient Bob at his smart home and his health conditions are continuously being monitored by health practitioners and his family members. 
%This entire test bed is deployed in 8 days.
The data generated from these devices is stored at their corresponding manufacturer's clouds (e.g., iHealth cloud, Wireless sensor cloud) and analyzed by their cloud services. At his smart home, motion sensors are attached at doors with 30$^{\circ}$ angle. If the angle of door is more than 30$^{\circ}$, the sensor sends the notification "open-door" to gateway device (e.g., Bob's smart phone). The generated data is pushed onto home automation's cloud, and his family members will also receive notification through IoT applications. Both motion sensors communicate through Ethernet tag manager, which has a connected light. When light turns ON it means that it is connected to the cloud based web service, when update light flashes it means that it is forwarding door sensor state (Open/Close) to the web service. Bob checks his body temperature, oxygen level, and blood pressure in every 3 hours, and connected health devices at object layer communicate to a gateway device (i.e., Bob's smart phone) which allows interaction with the upper layers of the architecture. During 15 days of monitoring, The range of Bob's body parameters are - SpO\textsubscript{2} (88\% - 98\%), blood pressure readings (110/65 mm Hg - 130/70 mm Hg), and  body temperature (98$^{\circ}$ - 103$^{\circ}$). When Bob's body temperature reached at 103$^{\circ}$, the health practitioners prescribed medicine and also informed his family members to watch him continuously. Bob was not able to sleep at night coughing, then his family members received notification from bedroom door sensor through home automation cloud service. They then communicated with him and asked him to check his oxygen level, which should not be less than 80\%. Figure \ref{fig:RMP} shows an overall sequential view of RPM at home along with data flow and interactions among entities and components of the system.
%There was no reference to Figure 10, so added here. 

This setup is an implementation of one of the smart communities' scenario, i.e., RPM. Similarly, other use case scenarios can be implemented using same technologies and solutions along with various IoT devices and utilizing our proposed implementation architecture. While we discuss an existing cloud-enabled IoT implementation architecture and relevant technologies, there are several challenges to scale it for enabling future smart communities with various such smart application scenarios. Therefore, enabling such highly distributed, autonomous, and interdisciplinary architecture in real-world environment demands future research to overcome current challenges and make smart communities a reality in the future.

%IoT devices are collecting his health parameters and are connected to the edge device Bob's phone through Bluetooth low energy (BLE) and Wi-Fi. 

%It enables smart health devices to exchange data/information with health practitioners and family members through remote patient monitoring applications, which allows to monitor the patient remotely by health practitioners/family members on AWS IoT platform.

%There is a requirement to setup greengrass on all smart devices, and all smart devices must use a supported platform (e.g. Architecture: Armv7l, OS: Linux; Distribution: Raspbian Buster).

\section{Open Challenges and Future Directions}
\label{sec:challenges}
\begin{figure*}[t]
\centering
\includegraphics[width=1.0\textwidth, height=.43\textheight]{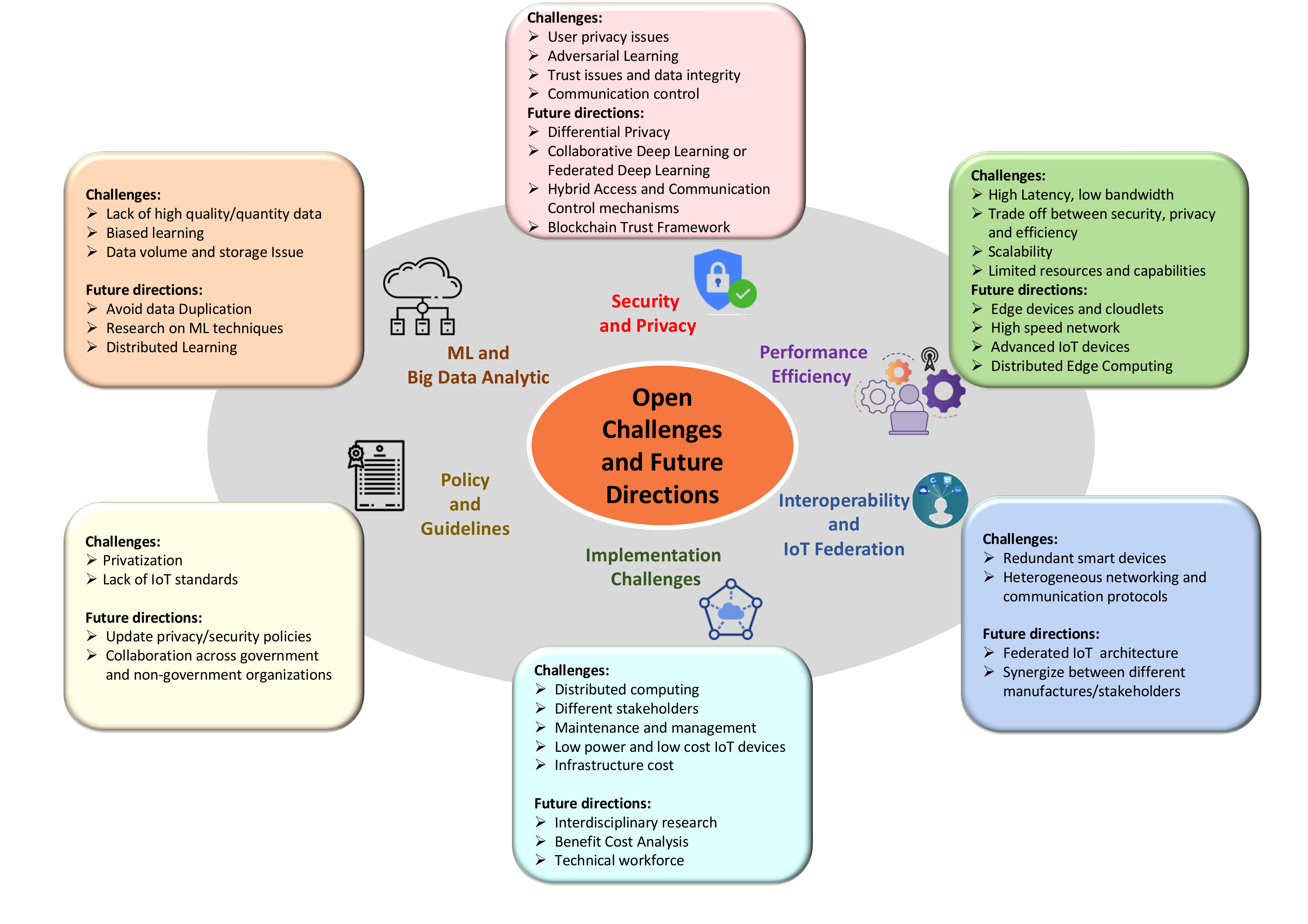}
\centering
\caption{Open Challenges and Future Directions in Smart Communities Deployment}
\label{fig:challenges}
\end{figure*}

The development of the proposed smart connected ecosystem requires to address several challenges and needs inter-disciplinary research from an integrated perspective involving different domains and stakeholders that we learned from proposed implementation of smart communities and implemented RPM system. In this section, we will discuss these challenges and required future research in detail with examples from each of the proposed scenarios, as illustrated in Figure \ref{fig:challenges}. %presents an overview of the key challenges together with future research directions. 

\subsection{Security and Privacy} One of the major challenges in the deployment of the smart infrastructures is the security and privacy concerns pertaining to IoT and CPS \cite{sontowski2020cyber} users, smart devices, data, and applications in different application domains like healthcare, smart home, supply chain management, transportation, and smart city. In health care industry, it is still a challenge to secure connected medical devices and ensure user privacy. 

In E-health scenario, for instance, a user visits smart testing booth for COVID-19 testing, and his/her data will transmit and store on smart hospital private cloud. Hospitals then share this data with state healthcare staff or city government for tracking and monitoring the user activities. To secure the identity of user and ensure privacy, differential privacy \cite{dwork2014algorithmic} and data masking techniques \cite{vasilomanolakis2015security} (pseudonymize \cite{kobsa2003privacy}, anonymize \cite{truta2006privacy}) can be used. However, there are limitations intrinsic to these solutions. In pseudonymize technique, data can be traced back into its original state with high risk of compromising user privacy, whereas it becomes impossible to return data into its original state in anonymize. It is critical to ensure user privacy while deploying IoT and data-driven applications for their wide-adoption in preventing, monitoring, and mitigating COVID-19. 

Secure authentication mechanisms including access control and communication control models are necessary for cloud-enabled IoT platforms to defend against unauthorized access and securing data, both at rest \cite{gupta2018attribute, gupta2017multi, gupta2017object,gupta2017poster, awaysheh2020next} and in motion. Several IoT access control models have been developed in the literature \cite{ouaddah2017access}, with cloud-assisted IoT access control models for AWS \cite{bhatt2017access1}, Google \cite{gupta2020access}, and Azure \cite{thakare2020parbac}. Traditional access control models are not adequate in addressing dynamic and evolving access control requirements in IoT. Attribute-Based Access Control (ABAC) \cite{hu2013guide, jin2012unified, gupta2016mathrm, gupta2018secure}, offers a flexible and dynamic access control model, which fits more into distributed IoT environments, such as smart home \cite{bhatt2019authorizations}, connected vehicles \cite{gupta2018authorization,gupta2019dynamic}, and wearable IoT devices \cite{bhatt2017access, bhatt2020poster}. 

In addition to access control, communications in terms of data flow between various components in cloud-enabled IoT platform need to be secured from unauthorized data access and modifications. Thus, Attribute-Based Communication Control (ABCC) \cite{bhatt2020abac} has been introduced. A novel lightweight authentication protocol \cite{kou2019lightweight} has been proposed to protect password, which is based on smart card and biometric identification. Moreover, for multi-cloud interaction and collaboration across several connected communities, there needs distributed computing enabled by dynamic trust frameworks. Further research on Blockchain technology \cite{tang2019iot} to enable a trusted framework among different cloud based platforms is necessary. Advanced access control techniques with a hybrid approach (including benefits of several access control models) need to be designed to give access to users at different levels. For example, in a smart connected community, users will have different levels of access on smart devices, data, and applications. With recent surge in contact tracing applications, user data privacy concerns are surfacing and will continue to arise in the future. Recent article\footnote{https://spectrum.ieee.org/telecom/security/tracking-covid19-with-the-iot-may-put-your-privacy-at-risk} have suggested the need to ensure privacy and develop privacy aware contact tracing solutions to balance public health and personal privacy of the users. A pertaining risk to these and other AI assisted system and applications is adversarial machine learning \cite{finlayson2019adversarial}, using which adversaries compromise user data and privacy. In order to protect the data sets, differential privacy \cite{dwork2014algorithmic, gautam2012improved} can be applied to add noise. Cloud based medical data storage and the upfront challenges have been extensively addressed in the literature \cite{hani2014development,nalinipriya2013extensive}. Study \cite{zeng2017end} conducted semi-structured interviews with fifteen people living in smart homes to learn about how they use their smart homes, and to understand their security and privacy concerns, expectations, and actions.

Various supervised ML algorithm are used for security such as classification of IoT data as "attack" or "normal", however these algorithms have not succeed to produce significant results due to distinct requirements of IoT devices including scalability, distribution, resource limitations, and low latency. This study \cite{rathore2018semi} introduced a fog-based attack detection framework and a ELM-based Semi-supervised Fuzzy C-Means (ESFCM) method, which are the extension of cloud computing, fog computing based framework enables attack detection at the network edge and ESFCM method uses a semi-supervised fuzzy c-means algorithm to handle the labeled data issue.

In future research, there are requirements to conduct interviews of health practitioners to understand the security and privacy concerns while developing the smart hospital, and need to apply similar approach involving community residents, infrastructure developers and stakeholders to develop other components of the smart connected ecosystem.  Privacy preserving deep learning \cite{mcdole2020analyzing, piplai2020knowledge} approaches such as collaborative deep learning or federated learning also need to be explored to train and deploy local models at the edge devices. A Blockchain-based secure DL \cite{rathore2019blockdeepnet} that combines DL and Blockchain to support secure collaborative DL in IoT. Collaborative DL approach is performed at the edge device level to avoid the third party, whereas Blockchain is  engaged to verify the confidentiality and integrity of collaborative DL in IoT.
\subsection{Performance Efficiency}
Within a connected ecosystem, users are constantly interacting with numerous smart devices and applications. One of main challenges in such an environment with billions of smart devices is performance efficiency and QoS. High latency, response time, packet loss, and low bandwidth are various parameters in a network system that affect the usability and feasibility of IoT systems. These devices generate large volumes of data which results in surge in data traffic that causes network congestion and latency. 

Due to high latency, it is not recommended to perform any computation on centralized cloud for critical healthcare services in an emergency situation. End-users and front-line workers prefer computation on edge devices and cloudlets to lower the wait and offer real-time analytics. This study \cite{li2019special} shows various approaches to improve accuracy and efficiency of applications.
Research and advancements in distributed edge computing technologies, low latency networking and communication protocols are essential to enable continued success of IoT-enabled smart environments. Data can be stored on multiple edge devices to improve processing efficiency and response time. Typically, high energy consumption required by mobile communications is a serious hurdle to large scale IoT deployments of use case scenarios. In near future, we expect to witness IoT applications expand in a world of 5G, with the proliferation of high speed networks with low latency applications, such as RPM, real-time Vehicle-to-Vehicle (V2V) and Vehicle-to-Infrastructure (V2I) communication. %Nonetheless, there is always a trade-off between security, privacy and efficiency. 
%As discussed above, adding noise or anonymity to the datasets also adds some performance overhead.  

\subsection{Interoperability and IoT Federation} Recently, there have been many smart devices introduced by various manufactures in the market. For example, there are several smart home assistants (e.g., Google Home, Alexa) that users can choose, however these devices use different communication protocols and also have connection with their corresponding cloud services. Interoperability among IoT devices has been an issue since the beginning of smart ecosystem deployments. The heterogeneity among billions of IoT devices adds to the challenge of developing smart connected communities. While in computer systems, TCP/IP is a standard suite of protocols that enable any system to communicate with other systems, such standards are still lacking in IoT domain. 

Another critical issue is how to reduce redundant smart devices. An effective solution to this problem would be to develop federated IoT and cloud architectures that can communicate with any type of sensors and devices. For instance, in different smart domains, a user received multiple wearable smart bracelets in smart communities. It will be challenging to keep track all alerts through their corresponding IoT applications. Therefore, there is a need to develop federated IoT architectures to enable federation across multiple smart devices and applications which can allow access and data communication across multiple IoT platforms. The research \cite{farris2017mifaas} suggested that IoT cloud provider allowed all devices belonging to the same private/public owner to participate in the federation process. However, many devices do not support other manufacturer's application platforms. For instance, in a smart home, all sensors and devices are connected through home assistant, and AI assisted application sends notification to home assistant. Due to interoperability or compatibility issue, home assistant could not receive the notifications from all devices. For future research, IoT software engineers, manufacturers and stakeholders should create synergies together to build smart devices and federated IoT architectures. 

\subsection{Policy and Guidelines} IoT is an emerging technology that is being adopted by several nations across the world. Countries\footnote{http://smartcities.gov.in/content/innerpage/guidelines.php} have different guidelines and policies for preserving privacy of users and their personal data \cite{weber2010internet,chatfield2019framework,tanczer2019united}, which are still evolving. 
Australian Privacy Principles (APP) \cite{caron2016internet} designed their own guidelines, procedures and standards to protect individual privacy associated with data collection through IoT. The U.S. Department of Transportation (USDOT) has similarly issued smart city challenge\footnote{https://www.transportation.gov/smartcity} and policy initiative in this direction. The U.S. government has committed approximately \$160 million\footnote{https://www.govtech.com/opinion/If-Only-One-US-City-Wins-the-Smart-City-Race-the-Whole-Nation-Loses.html} over the next five years to support smart city initiatives
smart connected communities with hospitals, homes, cities or transportation demand new constitutions and policies for preserving user data privacy and misuse of data and information. Such policies should maintain privacy of user and address another issue including accuracy, property and accessibility, together with infrastructure design to facilitate swift implementation of guidelines and public orders in COVID-19 and other similar situations. 
% Smart infrastructures are developing rapidly, however constitutions and policies are not updating and mapping with all of the new aspects introduced by IoT technologies. 

One reason for the lack of constitutions and policies may be because the IoT differs from other network technologies and there is a lack of specific IoT standards. The research on constitution and policy, including engagement on public policy development debates, and IoT standards is necessary to successfully integrate privacy, accuracy, property and accessibility solutions in the smart communities. To develop effective constitutional policies and standards, collaboration across governmental, non-governmental organizations and industry partners as cloud providers, IoT manufacturers, would be beneficial.  

\subsection{Implementation Challenges} 
Enabling a smart community requires thousands of low-power and low-cost embedded devices, which work together with large scale data analytics and applications. There are several implementation challenges involved in developing such large scale smart infrastructure. Fault tolerance and resilience are the challenges for reliable delivery of sensor data from smart devices to distributed cloud service. IoT applications need to fault tolerance since there can be various issues in applications or systems including face recognition, community infrastructure management, and emergency response in smart infrastructure. GEOgraphically Correlated Resilient Overlay Networks (GeoCRON) \cite{benson2016resilient} is developed to capture the localized nature of community IoT deployments in the context of small failures. Research \cite{muhammed2017enabling} proposed a new fault-tolerant routing technique for hierarchical sensor networks. Cloud service providers can be maximised their revenue using an optimal pricing scheme \cite{zhang2019resource}. Another challenge for constant running and managing these IoT devices is costs related to energy, communication, computation, infrastructure, and operation. There is generally a trade-off between benefit and cost for IoT applications \cite{van2018cost}, however in the scenario of COVID-19 pandemic, expected benefits (saving lives, economic growth) should outweigh the operational and deployment costs. For instance, to build RPM platform with various vendors - intel technologies, iHealth labs etc., which 
provide pre-validated platforms for faster and lower cost development. Combined implementation cost includes smart devices cost, servicing and monitoring cost for RPM platform, which can be range from \$275 – \$2300 per patient per year. As we setup test bed of RPM use case at home, the total cost of this test bed is \$310 and spent 8-10 days to setup this use case. 

Another challenge is to build prediction model by paying minimum cost, which may effect the accuracy of the model. This study \cite{skourletopoulos2017cost} proposed a game theoretical analysis to allocate more storage capacity in a cost-effective manner, which achieves to maximize the benefits. For the future directions, game theoretical approach can be used to analyze the smart infrastructures in terms of cost-benefit analysis. Furthermore, interdisciplinary research collaboration is inevitable to implement a smart connected ecosystem. There are several areas of research and engineering aspects, as machine-to-machine technology, artificial intelligence \cite{chukkapalli2020ontologies, chukkapalli2020smart}, ML, predictive analysis, security and privacy, and others need to be merged and collaborative research approach is necessary in implementing, deploying, and managing a smart connected ecosystem.

%Some private were strategic and forward-thinking in funding and partnering with sub-national governments in promoting the IoT use.
%Whereas, COVID-19 pandemic Smart communities can handle such specific issues social distancing measures, secure data collection, video surveillance and sanitizing the environment. However, the constitutions and policies are not updating regularly to deploy these connected infrastructures.

\subsection{ML and Big Data Analytic}
IoT generates tremendous amount of data, and this raw data is converted into valuable knowledge using AI and ML technologies. The "6V" (Value, Velocity, Volume, Variety, Variability, and Veracity) big data challenges for IoT applications are discussed in \cite{farhan2019internet, mohammadi2018deep}. The volume of data from IoT devices overwhelms storage capacities. There is not only storage issue, but data needs to be organized properly so that it can be retrieved and processed. Data duplication is an issue when an organization has multiple copies of the same data source. For example, a user has multiple wearable smart bracelets (smart hospital bracelet, smart grocery bracelet, and RFID antibody tag bracelet), these wearable devices will collect similar kind of data, which can create an issue of data duplication. 

ML based applications require a large amount of valuable data for correct prediction, however, complicated and insufficient data can be an issue to the accuracy of the learning and predictive models. In addition, ML approaches need further research and development to deal with such heterogeneous and constantly evolving sensory data inputs. For instance, a locality-based COVID-19 patient detection model can predict early symptoms, which will be based on the collaboration of smart nursing home data sets and smart child care data sets. The prediction model can be biased towards elderly people if the number of patients in smart nursing home are more than smart child care. To overcome this problem, both models can learn at their edge networks using collaborative deep learning \cite{dgupta2020}. Research on these open challenges will help early development and deployment of future smart communities.

\section{Conclusion}
\label{sec:conclusion}
In this paper, we propose future smart connected community scenarios, which are blueprints to develop smart and intelligent infrastructures against COVID-19 to stop pandemic outbreaks in the future. The autonomous operations with low human intervention in smart communities enable safe environment and enforce preventive measures for controlling the spread of infection in community. Data-driven and AI assisted applications facilitate increased testing, monitoring and tracking of COVID-19 patients, and help to enforce social distancing measures, predict possible infections based on symptoms and human activities, optimize the delivery of essential services and resources in a swift and efficient manner. 
This paper discussed different scenarios to reflect smart applications and connected ecosystem. Further, it proposed IoT enabled big data collection and analysis in different sectors including healthcare, home, supply chain management, transportation, environment, and city. A general implementation framework of smart communities is discussed on AWS cloud service. In addition, a proof of concept implementation of a RPM use case is presented. Furthermore, wider implementation of proposed smart connected communities faces many challenges, such as legislation and policy, deployment cost, time, security and privacy, which have also been discussed in the paper. We believe that our vision of future smart communities will ignite interdisciplinary research and development of connected ecosystem to prepare our world against future pandemic outbreaks. %such as E-health, smart home, smart grocery, smart locality, and smart city.

{
\footnotesize
\bibliographystyle{unsrt}
\bibliography{reference}

\begin{thebibliography}{100}

\bibitem{kabir2019state}
M~Humayun Kabir, Keshav Thapa, Jae-Young Yang, and Sung-Hyun Yang.
\newblock State-space based linear modeling for human activity recognition in
  smart space.
\newblock {\em INTELLIGENT AUTOMATION AND SOFT COMPUTING}, 25(4):673--681,
  2019.

\bibitem{bai2020chinese}
Li~Bai et~al.
\newblock Chinese experts’ consensus on the internet of things-aided
  diagnosis and treatment of coronavirus disease 2019 (covid-19).
\newblock {\em Clinical eHealth}, 3:7--15, 2020.

\bibitem{vaishya2020artificial}
Raju Vaishya, Mohd Javaid, Ibrahim~Haleem Khan, and Abid Haleem.
\newblock Artificial intelligence (ai) applications for covid-19 pandemic.
\newblock {\em Diabetes \& Metabolic Syndrome: Clinical Research \& Reviews},
  2020.

\bibitem{javaid2020industry}
Mohd Javaid, Abid Haleem, Raju Vaishya, Shashi Bahl, Rajiv Suman, and Abhishek
  Vaish.
\newblock Industry 4.0 technologies and their applications in fighting covid-19
  pandemic.
\newblock {\em Diabetes \& Metabolic Syndrome: Clinical Research \& Reviews},
  2020.

\bibitem{kumar2020role}
Krishna Kumar, Narendra Kumar, and Rachna Shah.
\newblock Role of iot to avoid spreading of covid-19.
\newblock {\em International Journal of Intelligent Networks}, 1:32--35, 2020.

\bibitem{vinod2020data}
Dasari~Naga Vinod and SRS Prabaharan.
\newblock Data science and the role of artificial intelligence in achieving the
  fast diagnosis of covid-19.
\newblock {\em Chaos, Solitons \& Fractals}, page 110182, 2020.

\bibitem{otoom2020iot}
Mwaffaq Otoom, Nesreen Otoum, Mohammad~A Alzubaidi, Yousef Etoom, and Rudaina
  Banihani.
\newblock An iot-based framework for early identification and monitoring of
  covid-19 cases.
\newblock {\em Biomedical Signal Processing and Control}, page 102149, 2020.

\bibitem{kumar2020drone}
Adarsh Kumar, Kriti Sharma, Harvinder Singh, Sagar~Gupta Naugriya,
  Sukhpal~Singh Gill, and Rajkumar Buyya.
\newblock A drone-based networked system and methods for combating coronavirus
  disease (covid-19) pandemic.
\newblock {\em Future Generation Computer Systems}, 2020.

\bibitem{lalmuanawma2020applications}
Samuel Lalmuanawma, Jamal Hussain, and Lalrinfela Chhakchhuak.
\newblock Applications of machine learning and artificial intelligence for
  covid-19 (sars-cov-2) pandemic: A review.
\newblock {\em Chaos, Solitons \& Fractals}, page 110059, 2020.

\bibitem{singh2020internet}
Ravi~Pratap Singh, Mohd Javaid, Abid Haleem, and Rajiv Suman.
\newblock Internet of things (iot) applications to fight against covid-19
  pandemic.
\newblock {\em Diabetes \& Metabolic Syndrome: Clinical Research \& Reviews},
  2020.

\bibitem{swayamsiddha2020application}
Swati Swayamsiddha and Chandana Mohanty.
\newblock Application of cognitive internet of medical things for covid-19
  pandemic.
\newblock {\em Diabetes \& Metabolic Syndrome: Clinical Research \& Reviews},
  2020.

\bibitem{elavarasan2020restructured}
Rajvikram~Madurai Elavarasan and Rishi Pugazhendhi.
\newblock Restructured society and environment: A review on potential
  technological strategies to control the covid-19 pandemic.
\newblock {\em Science of The Total Environment}, page 138858, 2020.

\bibitem{vafea2020emerging}
Maria~Tsikala Vafea et~al.
\newblock Emerging technologies for use in the study, diagnosis, and treatment
  of patients with covid-19.
\newblock {\em Cellular and molecular bioengineering}, pages 1--9, 2020.

\bibitem{alam2020internet}
Tanweer Alam.
\newblock Internet of things and blockchain-based framework for coronavirus
  (covid-19) disease.
\newblock {\em Available at SSRN 3660503}, 2020.

\bibitem{albahri2020role}
AS~Albahri, Rula~A Hamid, et~al.
\newblock Role of biological data mining and machine learning techniques in
  detecting and diagnosing the novel coronavirus (covid-19): A systematic
  review.
\newblock {\em Journal of Medical Systems}, 44(7), 2020.

\bibitem{stantchev2015smart}
Vladimir Stantchev, Ahmed Barnawi, Sarfaraz Ghulam, Johannes Schubert, and
  Gerrit Tamm.
\newblock Smart items, fog and cloud computing as enablers of servitization in
  healthcare.
\newblock {\em Sensors \& Transducers}, 185(2):121, 2015.

\bibitem{li2014enabling}
Yang Li, Li~Guo, and Yike Guo.
\newblock Enabling health monitoring as a service in the cloud.
\newblock In {\em 2014 IEEE/ACM 7th International Conference on Utility and
  Cloud Computing}, pages 127--136. IEEE, 2014.

\bibitem{fernandez2014opportunities}
Felipe Fernandez and George~C Pallis.
\newblock Opportunities and challenges of the internet of things for
  healthcare: Systems engineering perspective.
\newblock In {\em IEEE International Conference on Wireless Mobile
  Communication and Healthcare-Transforming Healthcare Through Innovations in
  Mobile and Wireless Technologies (MOBIHEALTH)}, pages 263--266, 2014.

\bibitem{gupta2020security}
Maanak Gupta, Mahmoud Abdelsalam, Sajad Khorsandroo, and Sudip Mittal.
\newblock {Security and Privacy in Smart Farming: Challenges and
  Opportunities}.
\newblock {\em IEEE Access}, 8:34564--34584, 2020.

\bibitem{hossain2016cloud}
M~Shamim Hossain and Ghulam Muhammad.
\newblock {Cloud-assisted industrial internet of things (IIoT)--enabled
  framework for health monitoring}.
\newblock {\em Computer Networks}, 101:192--202, 2016.

\bibitem{bhatt2017access}
Smriti Bhatt, Farhan Patwa, and Ravi Sandhu.
\newblock {An access control framework for cloud-enabled wearable internet of
  things}.
\newblock In {\em 2017 IEEE 3rd International Conference on Collaboration and
  Internet Computing (CIC)}, pages 328--338. IEEE, 2017.

\bibitem{yadav2019docker}
Anuj~Kumar Yadav, ML~Garg, et~al.
\newblock Docker containers versus virtual machine-based virtualization.
\newblock In {\em Emerging Technologies in Data Mining and Information
  Security}, pages 141--150. Springer, 2019.

\bibitem{satyanarayanan2009case}
Mahadev Satyanarayanan, Paramvir Bahl, Ram{\'o}n Caceres, and Nigel Davies.
\newblock The case for vm-based cloudlets in mobile computing.
\newblock {\em IEEE pervasive Computing}, 8(4):14--23, 2009.

\bibitem{nitti2015virtual}
Michele Nitti, Virginia Pilloni, Giuseppe Colistra, and Luigi Atzori.
\newblock The virtual object as a major element of the internet of things: a
  survey.
\newblock {\em IEEE Communications Surveys \& Tutorials}, 18(2):1228--1240,
  2015.

\bibitem{pandey2019application}
Vaibhav Pandey and Poonam Saini.
\newblock Application layer scheduling in cloud: Fundamentals, review and
  research directions.
\newblock {\em COMPUTER SYSTEMS SCIENCE AND ENGINEERING}, 34(6):357--376, 2019.

\bibitem{catarinucci2015iot}
Luca Catarinucci et~al.
\newblock {An IoT-aware architecture for smart healthcare systems}.
\newblock {\em IEEE Internet of Things Journal}, 2(6):515--526, 2015.

\bibitem{1213625}
{Sungmee Park} and S.~{Jayaraman}.
\newblock Enhancing the quality of life through wearable technology.
\newblock {\em IEEE Engineering in Medicine and Biology Magazine},
  22(3):41--48, 2003.

\bibitem{7070665}
L.~{Catarinucci}, D.~{de Donno}, L.~{Mainetti}, L.~{Palano}, L.~{Patrono},
  M.~L. {Stefanizzi}, and L.~{Tarricone}.
\newblock {An IoT-Aware Architecture for Smart Healthcare Systems}.
\newblock {\em IEEE Internet of Things Journal}, 2015.

\bibitem{1504802}
{Fabrice Axisa}, P.~M. {Schmitt}, C.~{Gehin}, G.~{Delhomme}, E.~{McAdams}, and
  A.~{Dittmar}.
\newblock {Flexible technologies and smart clothing for citizen medicine, home
  healthcare, and disease prevention}.
\newblock {\em IEEE Transactions on Information Technology in Biomedicine},
  9(3):325--336, 2005.

\bibitem{xu2019enabling}
Lei Xu, Chungen Xu, Zhongyi Liu, Yunling Wang, and Jianfeng Wang.
\newblock Enabling comparable search over encrypted data for iot with
  privacy-preserving.
\newblock {\em CMC-COMPUTERS MATERIALS \& CONTINUA}, 60(2):675--690, 2019.

\bibitem{deep2019survey}
Samundra Deep, Xi~Zheng, Chandan Karmakar, Dongjin Yu, Leonard~GC Hamey, and
  Jiong Jin.
\newblock A survey on anomalous behavior detection for elderly care using
  dense-sensing networks.
\newblock {\em IEEE Communications Surveys \& Tutorials}, 22(1):352--370, 2019.

\bibitem{li2019optimization}
Shen Li, Fang Liu, Jiayue Liang, Zhenhua Cai, and Zhiyao Liang.
\newblock Optimization of face recognition system based on azure iot edge.
\newblock {\em CMC-COMPUTERS MATERIALS \& CONTINUA}, 61(3):1377--1389, 2019.

\bibitem{li2019special}
Ying Li, Honghao Gao, and Yueshen Xu.
\newblock Special section on big data and service computing, 2019.

\bibitem{rathore2019blockseciotnet}
Shailendra Rathore et~al.
\newblock Blockseciotnet: Blockchain-based decentralized security architecture
  for iot network.
\newblock {\em Journal of Network and Computer Applications}, 143:167--177,
  2019.

\bibitem{bellavista2013convergence}
Paolo Bellavista, Giuseppe Cardone, Antonio Corradi, and Luca Foschini.
\newblock {Convergence of MANET and WSN in IoT urban scenarios}.
\newblock {\em IEEE Sensors Journal}, 13(10):3558--3567, 2013.

\bibitem{7165043}
D.~J. {Freitas}, T.~B. {Marcondes}, L.~H.~V. {Nakamura}, and R.~I.
  {Meneguette}.
\newblock A health smart home system to report incidents for disabled people.
\newblock In {\em 2015 International Conference on Distributed Computing in
  Sensor Systems}, pages 210--211, 2015.

\bibitem{6648859}
N.~{Hamlil} and M.~{Benabdellah}.
\newblock Study and implementation of a network point health smart home
  electrocardiographic.
\newblock In {\em International Conference on Advances in Biomedical
  Engineering}, pages 109--112, 2013.

\bibitem{7501687}
G.~{Sprint}, D.~{Cook}, R.~{Fritz}, and M.~{Schmitter-Edgecombe}.
\newblock {Detecting Health and Behavior Change by Analyzing Smart Home Sensor
  Data}.
\newblock In {\em 2016 IEEE International Conference on Smart Computing
  (SMARTCOMP)}, pages 1--3, 2016.

\bibitem{7574719}
M.~S. {Hossain}.
\newblock {Patient status monitoring for smart home healthcare}.
\newblock In {\em 2016 IEEE International Conference on Multimedia Expo
  Workshops (ICMEW)}, pages 1--6, 2016.

\bibitem{7206522}
P.~N. {Dawadi}, D.~J. {Cook}, and M.~{Schmitter-Edgecombe}.
\newblock {Automated Cognitive Health Assessment From Smart Home-Based Behavior
  Data}.
\newblock {\em IEEE Journal of Biomedical and Health Informatics}, 2016.

\bibitem{Okayode2020}
Olumide Kayode, Deepti Gupta, and Ali~Saman Tosun.
\newblock {Towards a Distributed Estimator in Smart Home Environment}.
\newblock {\em IEEE 6th World Forum on Internet of Things (WF-IoT)}, 2020.

\bibitem{gupta2020access}
Deepti Gupta, Smriti Bhatt, Maanak Gupta, Olumide Kayode, and Ali~Saman Tosun.
\newblock Access control model for google cloud iot.
\newblock In {\em 2020 IEEE 6th Intl Conference on Big Data Security on Cloud
  (BigDataSecurity), IEEE Intl Conference on High Performance and Smart
  Computing,(HPSC) and IEEE Intl Conference on Intelligent Data and Security
  (IDS)}, pages 198--208. IEEE, 2020.

\bibitem{Tekeoglu2015}
Ali Tekeoglu and Ali~\c{S}aman Tosun.
\newblock {A Closer Look into Privacy and Security of Chromecast Multimedia
  Cloud Communications}.
\newblock In {\em Multimedia Cloud Communication Workshop, Held in conjunction
  with INFOCOM 2015}, Hong Kong, China, April 2015.

\bibitem{Tekeoglu2015b}
Ali Tekeoglu and Ali~\c{S}aman Tosun.
\newblock {Investigating Security and Privacy of a Cloud-based Wireless IP
  Camera: NetCam}.
\newblock In {\em Workshop on Privacy, Security and Trust in Mobile and
  Wireless Systems (MobiPST 2015), Held in conjunction with ICCCN 2015}, Las
  Vegas, USA, August 2015.

\bibitem{Tekeoglu2016}
A.~Tekeoglu and A.~S. Tosun.
\newblock {A Testbed for Privacy and Security of IoT Devices}.
\newblock In {\em IEEE International Workshop on Data Science for Internet of
  Things, MASS 2016}, pages 1--6, Oct 2016.

\bibitem{Tekeoglu2017}
Ali Tekeoglu and Ali~\c{S}aman Tosun.
\newblock {An Experimental Framework for Investigating Security and Privacy of
  IoT Devices}.
\newblock In {\em International Conference on Intelligent, Secure, and
  Dependable Systems in Distributed and Cloud Environments}, Vancouver, CA,
  October 2017.

\bibitem{Kayode2019}
O.~{Kayode} and A.~S. {Tosun}.
\newblock {Analysis of IoT Traffic using HTTP Proxy}.
\newblock In {\em ICC 2019 - 2019 IEEE International Conference on
  Communications (ICC)}, pages 1--7, May 2019.

\bibitem{bhatt2017access1}
Smriti Bhatt, Farhan Patwa, and Ravi Sandhu.
\newblock Access control model for aws internet of things.
\newblock In {\em International Conference on Network and System Security},
  pages 721--736. Springer, 2017.

\bibitem{bhatt2019authorizations}
Smriti Bhatt, A~Tawalbeh Lo’ai, Pankaj Chhetri, and Paras Bhatt.
\newblock {Authorizations in cloud-based Internet of Things: current trends and
  use cases}.
\newblock In {\em 2019 Fourth International Conference on Fog and Mobile Edge
  Computing (FMEC)}, pages 241--246. IEEE, 2019.

\bibitem{kim2019dpn}
Dae-Young Kim, Se~Dong Min, and Seokhoon Kim.
\newblock A dpn (delegated proof of node) mechanism for secure data
  transmission in iot services.
\newblock {\em CMC Comput. Mater. Cont}, 60:1--14, 2019.

\bibitem{7750743}
{Yong Liang} et~al.
\newblock {A method to make accurate inventory of smart meters in multi-tags
  group-reading environment}.
\newblock In {\em 2016 IEEE International Conference on RFID Technology and
  Applications (RFID-TA)}, pages 123--128, 2016.

\bibitem{8780282}
J.~{Lin}, M.~{Zheng}, J.~{Chen}, K.~{He}, and E.~{Pan}.
\newblock {Smart Spare Part Inventory Management System with Sensor Data
  Updating}.
\newblock In {\em 2019 IEEE International Conference on Industrial Cyber
  Physical Systems (ICPS)}, pages 597--602, 2019.

\bibitem{7917105}
L.~{Zhang}, N.~{Alharbe}, and A.~S. {Atkins}.
\newblock {An IoT Application for Inventory Management with a Self-Adaptive
  Decision Model}.
\newblock In {\em 2016 IEEE International Conference on Internet of Things
  (iThings) and IEEE Green Computing and Communications (GreenCom) and IEEE
  Cyber, Physical and Social Computing (CPSCom) and IEEE Smart Data
  (SmartData)}, pages 317--322, 2016.

\bibitem{8939259}
W.~{Raad}, M.~{Bueno-Delgado}, M.~{Deriche}, and W.~{Suliman}.
\newblock {An IoT Based Inventory System for High Value Laboratory Equipment}.
\newblock In {\em 2019 Sixth International Conference on Internet of Things:
  Systems, Management and Security (IOTSMS)}, pages 314--319, 2019.

\bibitem{7932080}
R.~{Li}, T.~{Song}, N.~{Capurso}, J.~{Yu}, J.~{Couture}, and X.~{Cheng}.
\newblock {IoT Applications on Secure Smart Shopping System}.
\newblock {\em IEEE Internet of Things Journal}, 4(6):1945--1954, 2017.

\bibitem{8666395}
A.~{Oplas}, M.~H. {Rabago}, C.~L. {Tormes}, C.~L.~S. {Romana}, and
  R.~{Laviste}.
\newblock {Aeon: A Smart Medicine Delivery and Inventory System for Cebu City
  Government’s Long Life Medical Assistance Program}.
\newblock In {\em 2018 IEEE 10th International Conference on Humanoid,
  Nanotechnology, Information Technology,Communication and Control, Environment
  and Management (HNICEM)}, pages 1--6, 2018.

\bibitem{ding2013study}
Wen Ding.
\newblock {Study of smart warehouse management system based on the IoT}.
\newblock In {\em Intelligence computation and evolutionary computation}, pages
  203--207. Springer, 2013.

\bibitem{8743269}
S.~V. {Hove}, A.~{All}, and L.~D. {Marez}.
\newblock {Short on Time? Context-Aware Shopping Lists to the Rescue: an
  Experimental Evaluation of a Smart Shopping Cart}.
\newblock In {\em 2019 Eleventh International Conference on Quality of
  Multimedia Experience (QoMEX)}, pages 1--6, 2019.

\bibitem{9040750}
S.~{Mekruksavanich}.
\newblock {The Smart Shopping Basket Based on IoT Applications}.
\newblock In {\em 2019 IEEE 10th International Conference on Software
  Engineering and Service Science (ICSESS)}, pages 714--717, 2019.

\bibitem{raad2018sysmart}
Omar Raad, Majd Makdessi, Yazan Mohamad, and Issam Damaj.
\newblock Sysmart indoor services: A system of smart and connected
  supermarkets.
\newblock In {\em 2018 IEEE Canadian Conference on Electrical \& Computer
  Engineering (CCECE)}, pages 1--6. IEEE, 2018.

\bibitem{1607945}
G.~{Roussos}.
\newblock Enabling rfid in retail.
\newblock {\em Computer}, 39(3):25--30, 2006.

\bibitem{gupta2020secure}
Maanak Gupta, James Benson, Farhan Patwa, and Ravi Sandhu.
\newblock {Secure V2V and V2I Communication in Intelligent Transportation using
  Cloudlets}.
\newblock {\em IEEE Transactions on Services Computing}, 2020.

\bibitem{9187899}
M.~{Gupta}, F.~M. {Awaysheh}, J.~{Benson}, M.~{Al Azab}, F.~{Patwa}, and
  R.~{Sandhu}.
\newblock An attribute-based access control for cloud-enabled industrial smart
  vehicles.
\newblock {\em IEEE Transactions on Industrial Informatics}, pages 1--1, 2020.

\bibitem{gupta2019secure}
Maanak Gupta, James Benson, Farhan Patwa, and Ravi Sandhu.
\newblock Secure cloud assisted smart cars using dynamic groups and attribute
  based access control.
\newblock {\em arXiv preprint arXiv:1908.08112}, 2019.

\bibitem{gupta2020enabling}
Maanak Gupta et~al.
\newblock {Enabling and Enforcing Social Distancing Measures using Smart City
  and ITS Infrastructures: A COVID-19 Use Case}.
\newblock {\em arXiv preprint arXiv:2004.09246}, 2020.

\bibitem{6069711}
X.~{Li}, R.~{Lu}, X.~{Liang}, X.~{Shen}, J.~{Chen}, and X.~{Lin}.
\newblock {Smart community: an internet of things application}.
\newblock {\em IEEE Communications Magazine}, 49(11):68--75, 2011.

\bibitem{7406686}
Y.~{Sun}, H.~{Song}, A.~J. {Jara}, and R.~{Bie}.
\newblock {Internet of Things and Big Data Analytics for Smart and Connected
  Communities}.
\newblock {\em IEEE Access}, 4:766--773, 2016.

\bibitem{8355907}
P.~A. {Catherwood}, D.~{Steele}, M.~{Little}, S.~{Mccomb}, and J.~{Mclaughlin}.
\newblock {A Community-Based IoT Personalized Wireless Healthcare Solution
  Trial}.
\newblock {\em IEEE Journal of Translational Engineering in Health and
  Medicine}, 6:1--13, 2018.

\bibitem{7448556}
Y.~{Lee}, W.~{Hsiao}, C.~{Huang}, and S.~T. {Chou}.
\newblock {An integrated cloud-based smart home management system with
  community hierarchy}.
\newblock {\em IEEE Transactions on Consumer Electronics}, 62(1):1--9, 2016.

\bibitem{8656893}
K.~{Axel} and I.~S. {Khayal}.
\newblock {Modeling ‘Thriving Communities’ using a Systems Architecture to
  Improve Smart Cities Technology Approaches}.
\newblock In {\em IEEE International Smart Cities Conference (ISC2)}, pages
  1--2, 2018.

\bibitem{7384112}
H.~{Yan} and D.~{Long}.
\newblock {Research on Key Technology for Data Storage in Smart Community Based
  on Big Data}.
\newblock In {\em International Conference on Intelligent Transportation, Big
  Data and Smart City}, 2015.

\bibitem{mora2018use}
Olga~B Mora et~al.
\newblock A use case in cybersecurity based in blockchain to deal with the
  security and privacy of citizens and smart cities cyberinfrastructures.
\newblock In {\em 2018 IEEE International Smart Cities Conference (ISC2)},
  pages 1--4. IEEE, 2018.

\bibitem{6740844}
A.~{Zanella} et~al.
\newblock {Internet of Things for Smart Cities}.
\newblock {\em IEEE Internet of Things Journal}, 2014.

\bibitem{6702523}
J.~{Jin}, J.~{Gubbi}, S.~{Marusic}, and M.~{Palaniswami}.
\newblock {An Information Framework for Creating a Smart City Through Internet
  of Things}.
\newblock {\em IEEE Internet of Things Journal}, 1(2):112--121, 2014.

\bibitem{7721743}
M.~{Centenaro}, L.~{Vangelista}, A.~{Zanella}, and M.~{Zorzi}.
\newblock {Long-range communications in unlicensed bands: the rising stars in
  the IoT and smart city scenarios}.
\newblock {\em IEEE Wireless Communications}, 23(5):60--67, 2016.

\bibitem{6871673}
A.~{Solanas}, C.~{Patsakis}, M.~{Conti}, I.~S. {Vlachos}, V.~{Ramos}, et~al.
\newblock {Smart health: A context-aware health paradigm within smart cities}.
\newblock {\em IEEE Communications Magazine}, 52(8):74--81, 2014.

\bibitem{7876852}
H.~{Menouar}, I.~{Guvenc}, K.~{Akkaya}, A.~S. {Uluagac}, A.~{Kadri}, and
  A.~{Tuncer}.
\newblock {UAV-Enabled Intelligent Transportation Systems for the Smart City:
  Applications and Challenges}.
\newblock {\em IEEE Communications Magazine}, 55(3):22--28, 2017.

\bibitem{9084093}
X.~{Zheng}, L.~{Tian}, G.~{Luo}, and Z.~{Cai}.
\newblock {A Collaborative Mechanism for Private Data Publication in Smart
  Cities}.
\newblock {\em IEEE Internet of Things Journal}, pages 1--1, 2020.

\bibitem{9018282}
A.~{Meslin}, N.~{Rodriguez}, and M.~{Endler}.
\newblock {Scalable Mobile Sensing for Smart Cities: The MUSANet Experience}.
\newblock {\em IEEE Internet of Things Journal}, pages 1--1, 2020.

\bibitem{vengadeswaran2019core}
Vengadeswaran and Balasundaram.
\newblock Core-an optimal data placement strategy in hadoop for data intensive
  applications based on cohesion relation.
\newblock {\em COMPUTER SYSTEMS SCIENCE AND ENGINEERING}, 34(1):47--60, 2019.

\bibitem{554205}
James Llinas and David~L Hall.
\newblock An introduction to multi-sensor data fusion.
\newblock In {\em Proceedings of IEEE International Symposium on Circuits and
  Systems}, volume~6, pages 537--540, 1998.

\bibitem{250509}
P.~{Varaiya}.
\newblock {Smart cars on smart roads: problems of control}.
\newblock {\em IEEE Transactions on Automatic Control}, 38(2):195--207, 1993.

\bibitem{1306972}
J.~P. {Hubaux}, S.~{Capkun}, and {Jun Luo}.
\newblock {The security and privacy of smart vehicles}.
\newblock {\em IEEE Security Privacy}, 2(3):49--55, 2004.

\bibitem{791202}
D.~M. {Gavrila} and V.~{Philomin}.
\newblock {Real-time object detection for "smart" vehicles}.
\newblock In {\em Proceedings of the Seventh IEEE International Conference on
  Computer Vision}, volume~1, pages 87--93 vol.1, 1999.

\bibitem{4162483}
P.~{Falcone}, F.~{Borrelli}, J.~{Asgari}, H.~E. {Tseng}, and D.~{Hrovat}.
\newblock {Predictive Active Steering Control for Autonomous Vehicle Systems}.
\newblock {\em IEEE Transactions on Control Systems Technology},
  15(3):566--580, 2007.

\bibitem{7585053}
J.~{Funke}, M.~{Brown}, S.~M. {Erlien}, and J.~C. {Gerdes}.
\newblock {Collision Avoidance and Stabilization for Autonomous Vehicles in
  Emergency Scenarios}.
\newblock {\em IEEE Transactions on Control Systems Technology}, 2017.

\bibitem{7139555}
M.~{Kuderer}, S.~{Gulati}, and W.~{Burgard}.
\newblock {Learning driving styles for autonomous vehicles from demonstration}.
\newblock In {\em IEEE International Conference on Robotics and Automation
  (ICRA)}, pages 2641--2646, 2015.

\bibitem{gupta2019dynamic}
Maanak Gupta, James Benson, Farhan Patwa, and Ravi Sandhu.
\newblock Dynamic groups and attribute-based access control for next-generation
  smart cars.
\newblock In {\em Proceedings of the Ninth ACM Conference on Data and
  Application Security and Privacy}, pages 61--72, 2019.

\bibitem{gupta2020dynamic}
Maanak Gupta, James Benson, Farhan Patwa, and Ravinderpal Sandhu.
\newblock Dynamic groups and attribute-based access control for next-generation
  smart cars, September~10 2020.
\newblock US Patent App. 16/811,165.

\bibitem{sontowski2020cyber}
Sina Sontowski, Maanak Gupta, Sai Sree~Laya Chukkapalli, Mahmoud Abdelsalam,
  Sudip Mittal, Anupam Joshi, and Ravi Sandhu.
\newblock Cyber attacks on smart farming infrastructure.
\newblock {\em UMBC Student Collection}, 2020.

\bibitem{dwork2014algorithmic}
Cynthia Dwork, Aaron Roth, et~al.
\newblock {The algorithmic foundations of differential privacy}.
\newblock {\em Foundations and Trends{\textregistered} in Theoretical Computer
  Science}, 9(3--4):211--407, 2014.

\bibitem{vasilomanolakis2015security}
Emmanouil Vasilomanolakis et~al.
\newblock {On the security and privacy of Internet of Things architectures and
  systems}.
\newblock In {\em 2015 International Workshop on Secure Internet of Things
  (SIoT)}, pages 49--57. IEEE, 2015.

\bibitem{kobsa2003privacy}
Alfred Kobsa and J{\"o}rg Schreck.
\newblock Privacy through pseudonymity in user-adaptive systems.
\newblock {\em ACM Transactions on Internet Technology (TOIT)}, 3(2):149--183,
  2003.

\bibitem{truta2006privacy}
Traian~Marius Truta and Bindu Vinay.
\newblock {Privacy protection: p-sensitive k-anonymity property}.
\newblock In {\em 22nd International Conference on Data Engineering Workshops
  (ICDEW'06)}, pages 94--94. IEEE, 2006.

\bibitem{gupta2018attribute}
Maanak Gupta, Farhan Patwa, and Ravi Sandhu.
\newblock An attribute-based access control model for secure big data
  processing in hadoop ecosystem.
\newblock In {\em Proceedings of the Third ACM Workshop on Attribute-Based
  Access Control}, pages 13--24, 2018.

\bibitem{gupta2017multi}
Maanak Gupta et~al.
\newblock Multi-layer authorization framework for a representative hadoop
  ecosystem deployment.
\newblock In {\em Proceedings of the 22nd ACM on Symposium on Access Control
  Models and Technologies}, pages 183--190, 2017.

\bibitem{gupta2017object}
Maanak Gupta, Farhan Patwa, and Ravi Sandhu.
\newblock Object-tagged rbac model for the hadoop ecosystem.
\newblock In {\em IFIP Annual Conference on Data and Applications Security and
  Privacy}, pages 63--81. Springer, 2017.

\bibitem{gupta2017poster}
Maanak Gupta, Farhan Patwa, and Ravi Sandhu.
\newblock Poster: Access control model for the hadoop ecosystem.
\newblock In {\em Proceedings of the 22nd ACM on Symposium on Access Control
  Models and Technologies}, pages 125--127, 2017.

\bibitem{awaysheh2020next}
Feras~M Awaysheh, Mamoun Alazab, Maanak Gupta, Tom{\'a}s~F Pena, and Jos{\'e}~C
  Cabaleiro.
\newblock Next-generation big data federation access control: A reference
  model.
\newblock {\em Future Generation Computer Systems}, 2020.

\bibitem{ouaddah2017access}
Aafaf Ouaddah, Hajar Mousannif, Anas~Abou Elkalam, and Abdellah~Ait Ouahman.
\newblock {Access control in the Internet of Things: Big challenges and new
  opportunities}.
\newblock {\em Computer Networks}, 112:237--262, 2017.

\bibitem{thakare2020parbac}
Abhijeet Thakare, Euijong Lee, Ajay Kumar, Valmik~B Nikam, and Young-Gab Kim.
\newblock {{PARBAC: Priority-Attribute-Based RBAC Model for Azure IoT Cloud}}.
\newblock {\em IEEE Internet of Things Journal}, 7(4):2890--2900, 2020.

\bibitem{hu2013guide}
Vincent~C Hu et~al.
\newblock {Guide to attribute based access control (abac) definition and
  considerations (draft)}.
\newblock {\em NIST special publication}, 800(162), 2013.

\bibitem{jin2012unified}
Xin Jin, Ram Krishnan, and Ravi Sandhu.
\newblock {A unified attribute-based access control model covering DAC, MAC and
  RBAC}.
\newblock In {\em IFIP Annual Conference on Data and Applications Security and
  Privacy}, pages 41--55. Springer, 2012.

\bibitem{gupta2016mathrm}
Maanak Gupta and Ravi Sandhu.
\newblock The $\mathrm{GURA_G}$ administrative model for user and group
  attribute assignment.
\newblock In {\em International Conference on Network and System Security},
  pages 318--332. Springer, 2016.

\bibitem{gupta2018secure}
Maanak Gupta.
\newblock {\em Secure Cloud Assisted Smart Cars and Big Data: Access Control
  Models and Implementation}.
\newblock PhD thesis, The University of Texas at San Antonio, 2018.

\bibitem{gupta2018authorization}
Maanak Gupta and Ravi Sandhu.
\newblock Authorization framework for secure cloud assisted connected cars and
  vehicular internet of things.
\newblock In {\em Proceedings of the 23nd ACM on Symposium on Access Control
  Models and Technologies}, pages 193--204, 2018.

\bibitem{bhatt2020poster}
Paras Bhatt, Smriti Bhatt, and Myung Ko.
\newblock {{Poster: IoT SENTINEL-An ABAC Approach Against Cyber-Warfare In
  Organizations}}.
\newblock In {\em Proceedings of the 25th ACM Symposium on Access Control
  Models and Technologies}, pages 223--225, 2020.

\bibitem{bhatt2020abac}
Smriti Bhatt and Ravi Sandhu.
\newblock {{ABAC-CC: Attribute-Based Access Control and Communication Control
  for Internet of Things}}.
\newblock In {\em Proceedings of the 25th ACM Symposium on Access Control
  Models and Technologies}, pages 203--212, 2020.

\bibitem{kou2019lightweight}
Liang Kou, Yiqi Shi, Liguo Zhang, Duo Liu, and Qing Yang.
\newblock A lightweight three-factor user authentication protocol for the
  information perception of iot.
\newblock {\em Computers Materials \& Continua}, 58(2):545--565, 2019.

\bibitem{tang2019iot}
Bo~Tang, Hongjuan Kang, Jingwen Fan, Qi~Li, and Ravi Sandhu.
\newblock {Iot passport: a blockchain-based trust framework for collaborative
  internet-of-things}.
\newblock In {\em Proceedings of the 24th ACM Symposium on Access Control
  Models and Technologies}, pages 83--92, 2019.

\bibitem{finlayson2019adversarial}
Samuel~G Finlayson, John~D Bowers, Joichi Ito, Jonathan~L Zittrain, Andrew~L
  Beam, and Isaac~S Kohane.
\newblock {Adversarial attacks on medical machine learning}.
\newblock {\em Science}, 363(6433):1287--1289, 2019.

\bibitem{gautam2012improved}
Ajeet~Kumar Gautam, Vidushi Sharma, Shiv Prakash, and Maanak Gupta.
\newblock Improved hybrid intrusion detection system (hids): mitigating false
  alarm in cloud computing.
\newblock {\em BL Joshi}, page 101, 2012.

\bibitem{hani2014development}
Ahmad Fadzil~M Hani et~al.
\newblock {Development of private cloud storage for medical image research
  data}.
\newblock In {\em IEEE International Conference on Computer and Information
  Sciences}, pages 1--6, 2014.

\bibitem{nalinipriya2013extensive}
G~Nalinipriya and R~Aswin Kumar.
\newblock {Extensive medical data storage with prominent symmetric algorithms
  on cloud-a protected framework}.
\newblock In {\em INTERNATIONAL CONFERENCE ON SMART STRUCTURES AND
  SYSTEMS-ICSSS'13}, pages 171--177. IEEE, 2013.

\bibitem{zeng2017end}
Eric Zeng, Shrirang Mare, and Franziska Roesner.
\newblock {End user security and privacy concerns with smart homes}.
\newblock In {\em Thirteenth Symposium on Usable Privacy and Security
  ($\{$SOUPS$\}$ 2017)}, pages 65--80, 2017.

\bibitem{rathore2018semi}
Shailendra Rathore and Jong~Hyuk Park.
\newblock Semi-supervised learning based distributed attack detection framework
  for iot.
\newblock {\em Applied Soft Computing}, 72:79--89, 2018.

\bibitem{mcdole2020analyzing}
Andrew McDole, Mahmoud Abdelsalam, Maanak Gupta, and Sudip Mittal.
\newblock {Analyzing CNN Based Behavioural Malware Detection Techniques on
  Cloud IaaS}.
\newblock {\em arXiv preprint arXiv:2002.06383}, 2020.

\bibitem{piplai2020knowledge}
Aritran Piplai, Sudip Mittal, Mahmoud Abdelsalam, Maanak Gupta, Anupam Joshi,
  Tim Finin, et~al.
\newblock Knowledge enrichment by fusing representations for malware threat
  intelligence and behavior.
\newblock 2020.

\bibitem{rathore2019blockdeepnet}
Shailendra Rathore et~al.
\newblock Blockdeepnet: a blockchain-based secure deep learning for iot
  network.
\newblock {\em Sustainability}, 11(14):3974, 2019.

\bibitem{farris2017mifaas}
Ivan Farris, Leonardo Militano, Michele Nitti, Luigi Atzori, and Antonio Iera.
\newblock {MIFaaS: A mobile-IoT-federation-as-a-service model for dynamic
  cooperation of IoT cloud providers}.
\newblock {\em Future Generation Computer Systems}, 70:126--137, 2017.

\bibitem{weber2010internet}
Rolf~H Weber.
\newblock {Internet of Things-New security and privacy challenges}.
\newblock {\em Computer law \& security review}, 26(1):23--30, 2010.

\bibitem{chatfield2019framework}
Akemi~Takeoka Chatfield and Christopher~G Reddick.
\newblock {A framework for Internet of Things-enabled smart government: A case
  of IoT cybersecurity policies and use cases in US federal government}.
\newblock {\em Government Information Quarterly}, 36(2):346--357, 2019.

\bibitem{tanczer2019united}
Leonie Tanczer, Irina Brass, Miles Elsden, Madeline Carr, and Jason~J
  Blackstock.
\newblock {The United Kingdom's Emerging Internet of Things (IoT) Policy
  Landscape}.
\newblock {\em Tanczer, LM, Brass, I., Elsden, M., Carr, M., \& Blackstock,
  J.(2019). The United Kingdom’s Emerging Internet of Things (IoT) Policy
  Landscape. In R. Ellis \& V. Mohan (Eds.), Rewired: Cybersecurity
  Governance}, pages 37--56, 2019.

\bibitem{caron2016internet}
Xavier Caron et~al.
\newblock {The Internet of Things (IoT) and its impact on individual privacy:
  An Australian perspective}.
\newblock {\em Computer Law \& Security Review}, 32(1):4--15, 2016.

\bibitem{benson2016resilient}
Kyle~E Benson, Qing Han, Kyungbaek Kim, Phu Nguyen, and Nalini
  Venkatasubramanian.
\newblock {Resilient overlays for IoT-based community infrastructure
  communications}.
\newblock In {\em IEEE International Conference on Internet-of-Things Design
  and Implementation (IoTDI)}, 2016.

\bibitem{muhammed2017enabling}
Thaha Muhammed, Rashid Mehmood, and Aiiad Albeshri.
\newblock {Enabling reliable and resilient IoT based smart city applications}.
\newblock In {\em International Conference on Smart Cities, Infrastructure,
  Technologies and Applications}, pages 169--184. Springer, 2017.

\bibitem{zhang2019resource}
Haiyang Zhang, Guolong Chen, and Xianwei Li.
\newblock Resource management in cloud computing with optimal pricing policies.
\newblock {\em COMPUTER SYSTEMS SCIENCE AND ENGINEERING}, 34(4):249--254, 2019.

\bibitem{van2018cost}
M~Jeroen Van Der~Donckt, Danny Weyns, M~Usman Iftikhar, and Ritesh~Kumar Singh.
\newblock {Cost-Benefit Analysis at Runtime for Self-adaptive Systems Applied
  to an Internet of Things Application.}
\newblock In {\em ENASE}, pages 478--490, 2018.

\bibitem{skourletopoulos2017cost}
Georgios Skourletopoulos et~al.
\newblock {Cost-benefit analysis game for efficient storage allocation in
  cloud-centric internet of things systems: a game theoretic perspective}.
\newblock In {\em IFIP/IEEE Symposium on Integrated Network and Service
  Management (IM)}, pages 1149--1154, 2017.

\bibitem{chukkapalli2020ontologies}
Sai Sree~Laya Chukkapalli, Sudip Mittal, Maanak Gupta, Mahmoud Abdelsalam,
  Anupam Joshi, Ravi Sandhu, Karuna~Pande Joshi, et~al.
\newblock Ontologies and artificial intelligence systems for the cooperative
  smart farming ecosystem.
\newblock {\em IEEE Access}, 8:164045--164064, 2020.

\bibitem{chukkapalli2020smart}
Sai Sree~Laya Chukkapalli, Aritran Piplai, Sudip Mittal, Maanak Gupta, Anupam
  Joshi, et~al.
\newblock A smart-farming ontology for attribute based access control.
\newblock In {\em 6th IEEE International Conference on Big Data Security on
  Cloud (BigDataSecurity 2020)}, 2020.

\bibitem{farhan2019internet}
Laith Farhan and Rupak Kharel.
\newblock {Internet of things: Vision, future directions and opportunities}.
\newblock In {\em Modern Sensing Technologies}, pages 331--347. Springer, 2019.

\bibitem{mohammadi2018deep}
Mehdi Mohammadi, Ala Al-Fuqaha, Sameh Sorour, and Mohsen Guizani.
\newblock {{Deep learning for IoT big data and streaming analytics: A survey}}.
\newblock {\em IEEE Communications Surveys \& Tutorials}, 2018.

\bibitem{dgupta2020}
Deepti Gupta, Olumide Kayode, Smriti Bhatt, Maanak Gupta, and Ali~Saman Tosun.
\newblock {Learner’s Dilemma: IoT Devices Training Strategies in
  Collaborative Deep Learning}.
\newblock {\em IEEE 6th World Forum on Internet of Things (WF-IoT)}, 2020.

\end{thebibliography}
}

\end{document}